\pdfoutput=1
\documentclass[%
preprint,
preprintnumbers,
nofootinbib,
amsmath,amssymb,
pra,
]{revtex4-2}

\usepackage{graphicx}
\usepackage{dcolumn}
\usepackage{bm}
\usepackage{nicematrix}
\usepackage{upgreek}
\usepackage{xcolor}
\usepackage{todonotes}
\usepackage{IEEEtrantools}
\usepackage{dsfont}
\usepackage{siunitx}
\usepackage{mleftright}
\usepackage{mathtools}
\usepackage{booktabs}
\usepackage{multirow}
\usepackage{xspace}
\usepackage[caption=false]{subfig}
\usepackage{placeins}
\usepackage{orcidlink}

\usepackage{diffcoeff}
\difdef{l}{}{op-symbol = \mathrm{d}}

\PassOptionsToPackage{breaklinks}{hyperref}
\usepackage{hyperref}
\hypersetup{
	colorlinks = true,
	citecolor={blue},
	linkcolor={red},
	allbordercolors = {white},
	urlcolor={blue},
	breaklinks=true
}
\usepackage{cleveref}

\newcommand{\upe}{\mathrm{e}}
\newcommand{\upi}{\mathrm{i}}
\newcommand{\upA}{\mathrm{A}}
\newcommand{\upB}{\mathrm{B}}
\newcommand{\upp}{\mathrm{p}}
\newcommand{\ups}{\mathrm{s}}
\newcommand{\upl}{\mathrm{l}}
\newcommand{\upc}{\mathrm{c}}
\newcommand{\upT}{\mathrm{T}}
\newcommand{\upI}{\mathrm{I}}
\newcommand{\upII}{\mathrm{II}}

\newcommand{\Id}{\mathds{1}}

\newcommand{\PS}{R}
\newcommand{\BS}{B}

\newcommand{\vac}{\mathrm{vac}}

\newcommand{\cc}{\mathrm{c.c.}}

\DeclareMathOperator{\diag}{diag}

\DeclareMathOperator{\Tr}{Tr}

\DeclareMathOperator{\rect}{rect}
\DeclareMathOperator{\sinc}{sinc}

\newcommand*{\mqty}[1]{
	\begin{pmatrix}
		#1
	\end{pmatrix}
}

\DeclarePairedDelimiter\abs{\lvert}{\rvert}
\DeclarePairedDelimiter\norm{\lVert}{\rVert}

\DeclarePairedDelimiter\ket{\lvert}{\rangle}
\DeclarePairedDelimiterX\innerp[2]{\langle}{\rangle}{#1\,\delimsize\vert\,\mathopen{}#2}
\DeclarePairedDelimiterX\outerp[2]{\lvert}{\rvert}{\,#1\delimsize\rangle\delimsize\langle\mathopen{}#2\,}
\DeclarePairedDelimiterX\braket[3]{\langle}{\rangle}{#1\,\delimsize\vert\,\mathopen{}#2\,\delimsize\vert\,\mathopen{}#3}

\newcommand*{\refcite}[1]{ref.~\onlinecite{#1}}

\newcommand{\IEEElabel}[1]{\addtocounter{equation}{-1}\refstepcounter{equation}\label{#1}}

\newcommand{\typezero}{type\protect\nobreakdash-0\xspace} \newcommand{\Typezero}{Type\protect\nobreakdash-0\xspace}   \newcommand{\typezeroI}{type\protect\nobreakdash-0/I\xspace}

\newcommand{\typeII}{type\protect\nobreakdash-II\xspace} 
\newcommand{\TypeII}{Type\protect\nobreakdash-II\xspace}

\begin{document}
\count\footins = 1000

\title{Frequency-Resolved Simulations of Highly Entangled Biphoton States:\\Beyond the Single-Pair Approximation. II. Application to Entanglement-based Quantum Key Distribution}

\author{\orcidlink{0009-0009-9005-702X}\,Philipp Kleinpa\ss}
\email{philipp.kleinpass@dlr.de}
\affiliation{%
 Institute of Communications and Navigation, German Aerospace Center (DLR)
 }%
\affiliation{%
 Institute of Applied Physics, Technical University of Darmstadt%\\
}%

\author{\orcidlink{0000-0001-8114-1785}\,Thomas Walther and \orcidlink{0000-0002-1046-6176}\,Erik Fitzke}
\email{erik.fitzke@physik.tu-darmstadt.de}
\affiliation{
 Institute of Applied Physics, Technical University of Darmstadt%\\
}%

\date{\today}% It is always \today, today,

\begin{abstract}
We present time- and frequency-resolved simulations of quantum key distribution~(QKD) systems employing highly entangled biphoton quantum states. Our simulations are based on expansions of the covariance matrix and photon detection probabilities of biphoton states in terms of increasing orders of the joint spectral amplitude that were introduced in the first part of this series. Employing these expansions allows us to efficiently evaluate the impact of multi-pair events on the performance of the QKD systems while systematically taking into account effects from the photon spectra and many relevant imperfections of the setup. The results are shown to be in agreement with corresponding measurements of the key rates and quantum bit error rates.
\end{abstract}

\maketitle

\section{Introduction}
Entanglement has become a vital resource for many applications throughout different fields in quantum information~\cite{Ekert_1991, BBM_1992, Bennett_1996, Tittel_2000}. Entangled biphoton states can be conveniently generated by exciting quantum optical non-linear processes such as spontaneous parametric down-conversion~(SPDC) or spontaneous four-wave mixing~(SFWM)~\cite{Grice_1997, Law_2000, Yang_2008, Helt_2010, Christ_2011, Euler_2021}.
The necessity to perform accurate simulations of systems employing entangled biphoton states grows with the increasing amount of applications that require the distribution of entanglement. In most applications and experiments, states consisting of a single photon pair are desired, while the aforementioned non-linear optical processes always exhibit a non-zero probability of producing multiple pairs. This leads to the potential presence of uncorrelated photons from different pairs which may degrade the performance of the corresponding application.

One of the most popular examples of such an application is entanglement-based quantum key distribution~(QKD). QKD is a method to distribute a secret random bit string (keys) between a pair of users that can subsequently be used for classic data encryption. The security of the key establishment is based on fundamental quantum-physical principles. In entanglement-based QKD, multi-photon events lead to uncorrelated detections and thereby increase the quantum bit error rate~(QBER), ultimately limiting the achievable secure key rate when increasing the power driving the non-linear optical process. Thus, reliably predicting and assessing the performance of such systems requires the inclusion of multi-pair events.

To accurately model multi-pair generation processes, the phase-space formalism of Gaussian states can be employed~\cite{Ma_1990, Weedbrook_2012, Olivares_2012, Adesso_2014, Takeoka_2015, Fitzke_2023}. This formalism takes into account the entire photon statistics, including all orders of multi-photon effects. Those statistics, however, depend on the joint spectral amplitude (JSA) of the process, hence the spectral or temporal degree of freedom needs to be included. For highly entangled states, this leads to significant computational demands when methods based on the numerical Schmidt-decomposition of the JSA are applied directly~\cite{Mauerer_2009_Thesis, Lamata_2005}. For example, a convenient way to approximate the Schmidt decomposition is to discretize the spectrum on a sufficiently fine grid and to perform a singular value decomposition of the resulting square matrix~\cite{Mauerer_2009_Thesis}, for which the operation count scales approximately with the third power of the matrix dimension~\cite{Trefethen_1997}.

Due to the increased numerical efforts, models considering the temporal and spectral degrees of freedom often neglect multi-pair effects~\cite{Keller_1997, Grice_1997, Law_2000, Mikhailova_2008, Lee_2014, Liu_2020, Phehlukwayo_2020, Dorfman_2021}. 
To enable efficient simulations taking both spectral effects and multi-pair effects into account, we presented methods to systematically approximate the covariance matrix and detection statistics for highly entangled biphoton states in the first part of this two-part series, \refcite{Kleinpass_2024_partI}.

In the second part at hand, we apply these methods to perform time- and frequency-resolved simulations of an entanglement-based QKD system for four users that was presented in refs.~\cite{Fitzke_2022_4Party, Dolejsky_2023}, featuring a time-bin encoded version of the BBM92~\cite{BBM_1992} protocol. To efficiently simulate the state after splitting the photons by wavelength-division demultiplexing, we introduce a restriction of the JSA to only the relevant frequency components. Combining the covariance formalism with continuous degrees of freedom for time and frequency allows us to consider essentially all relevant effects in the setup such as multi-photon-pair emission, the real spectra of the photon pairs and demultiplexers, chromatic dispersion in the transmission fibers, imperfect two-photon interference, and realistic single-photon detectors exhibiting dark counts, dead times and afterpulses. Indeed, the simulation results are shown to be in excellent agreement with measurements.

The remainder of the article is structured as follows. In \cref{sec:Review_partI}, some results from the first part of the series, \refcite{Kleinpass_2024_partI}, are briefly reviewed for reference. 
In \cref{sec:SetupAndTransformations}, the QKD systems and details of the encoding of the quantum bits are presented. The setup may be operated either using orthogonal polarized photon pairs generated by \typeII SPDC, which are separated by their polarization, or using parallel polarized photon pairs generated by \typezero SPDC, which are separated by wavelength-division demultiplexing. \Cref{sec:Simulation_Model_and_Results} discusses the details of the simulation model and simulation results, starting from the model of the photon pair generation and their separation over the transmission through optical fibers to their detection in single-photon detectors. Finally, effects limiting the performance of the QKD systems such as chromatic dispersion in optical fibers are examined.

\section{Essentials from Part I: Continuous-Mode Covariance Formalism for Entangled Biphoton States}
\label{sec:Review_partI}

A common approach~\cite{Keller_1997, Grice_1997, Law_2000, Mikhailova_2008, Lee_2014, Liu_2020, Phehlukwayo_2020, Dorfman_2021} to model biphoton states with angular frequencies $\omega_\ups$ and $\omega_\upi$ generated by SPDC or SFWM is to approximate them up to the first contributing order in the photon creation operators $\hat{a}^\dag$:  
\begin{equation}\label{eq:SPDC_linear}
    \ket{\psi} \propto \mleft( \ket{0} + \frac{C}{2} \int \dl\omega_\ups \dl\omega_\upi \psi(\omega_\ups,\omega_\upi) \hat{a}_\nu^\dag(\omega_\ups) \hat{a}_\rho^\dag(\omega_\upi) \ket{0} \mright)\,.
\end{equation}
This state consists of a superposition between no pair being generated and the generation of exactly one photon pair, with the coefficient $C$ depending on the strength of the non-linear interaction and pump field. The coefficients $\nu, \rho$ label additional degrees of freedom, specifically the polarization of the generated photons, where $\nu=\rho$ for \typezeroI processes and $\nu\neq\rho$ for \typeII processes.
The possibility of generating multiple pairs of photons simultaneously is neglected. 

The covariance formalism on the other hand allows to capture such effects. Here the biphoton states are described by their covariance matrix  ${\bm{\gamma} = \exp(2 \bm{Z})}$, where
\begin{IEEEeqnarray}{c"c}\label{eq:Z_definition}
    \bm{Z}^{(\upI)} = C \mqty{ 0 & \bm{\psi} \\ \bm{\psi}^\dag & 0 } \,, &
    \bm{Z}^{(\upII)} = \frac{C}{2} 
    \mqty{
        0 & 0 & 0 & \bm{\psi}\\
        0 & 0 & \bm{\psi}^\upT & 0\\
        0 & \bm{\psi}^* & 0 & 0\\
        \bm{\psi}^\dag & 0 & 0 & 0
    } \,, \IEEEnonumber\\
\end{IEEEeqnarray}
for \typezeroI and \typeII processes, respectively and $\bm{\psi}$ is the integral operator whose kernel is the JSA $\psi(\omega_\ups,\omega_\upi)$. Employing the Schmidt decomposition, the JSA can be written as ${\bm{\psi} = \bm{U} \bm{\Sigma} \bm{V}^\dag}$, with the diagonal matrix $\bm{\Sigma}$ containing the Schmidt coefficients. The corresponding squeezing parameters are given by
\begin{IEEEeqnarray}{c"c}\label{eq:SqueezingParameters_Definition}
    \bm{\sigma}^{(\upI)} = 2C \bm{\Sigma} \,,&
    \bm{\sigma}^{(\upII)} = C \bm{\Sigma} \,,
    \IEEEeqnarraynumspace
\end{IEEEeqnarray}
leading to 
\begin{subequations}\label{eq:Covariance_Type0TypeII}
    \begin{IEEEeqnarray}{rCl}
        \bm{\gamma}^{(\upI)} &=& 
        \mqty{
            \bm{U} \cosh (\bm{\sigma}^{(\upI)}) \bm{U}^\dag &
            \bm{U} \sinh(\bm{\sigma}^{(\upI)}) \bm{V}^\dag\\
            \bm{V} \sinh(\bm{\sigma}^{(\upI)}) \bm{U}^\dag &
            \bm{V} \cosh(\bm{\sigma}^{(\upI)}) \bm{V}^\dag
        } \,, \IEEElabel{eq:Covariance_Type0} \IEEEeqnarraynumspace\\[5pt]
        \bm{\gamma}^{(\upII)} &=&
        \mqty{
            \bm{U} \cosh(\bm{\sigma}^{(\upII)}) \bm{U}^\dag & \bm{U} \sinh(\bm{\sigma}^{(\upII)}) \bm{V}^\dag\\
            \bm{V} \sinh(\bm{\sigma}^{(\upII)}) \bm{U}^\dag & \bm{V} \cosh(\bm{\sigma}^{(\upII)}) \bm{V}^\dag
        } \bigoplus \text{c.c.} \,. \IEEEnonumber\IEEElabel{eq:Covariance_TypeII}\\
    \end{IEEEeqnarray}
\end{subequations}

The full multivariate probability distribution for the detection of ${\bm{n} = (n_1, n_2, \dots n_D)}$ photons in detectors ${d = 1, \dots, D}$ can be calculated by repeatedly differentiating the probability-generating function
\begin{equation}\label{eq:GeneratingFunctionGaussian}
    G(\bm{w}) = \frac{\exp\mleft( -\bm{\alpha}^\dag (\mathds{1} + \bm{W} \bm{\varGamma})^{-1} \bm{\alpha}/2 \mright)}{\sqrt{\det(\mathds{1} + \bm{W} \bm{\varGamma})}} 
\end{equation} of the photon statistics~\cite{Thomas_2021, Nauth_2022, Fitzke_2023}
\begin{equation}\label{eq:PND_Formuala}
    P(\bm n) = \bm{\mathcal{D}_{\bm{w}}^{\bm{n}}}\, G(\bm{w}) = \mleft(\prod_{d=1}^D\frac{1}{n_d!} \diffp[n_d]{}{-w_d}\mright) G(\bm{w}) \bigg\vert_{\bm{w}=\bm{1}}\,,
\end{equation}
with the \emph{renormalized covariance}
\begin{equation}\label{eq:RenormalizedCovariance_Definition}
    \bm{\varGamma} = \frac{\bm{\gamma} - \Id}{2}
\end{equation}
and ${\bm W = \diag(\bm w)^{\oplus 2}}$.

For highly entangled states, the renormalized covariance can be approximated by an $N$-th order series expansion:
\begin{equation}\label{eq:RenormalizedCovariance_SeriesExpansion}
    \bm{\varGamma} \approx\bm{\varGamma}_N  =  \sum_{n=1}^N \frac{(2\bm{Z})^n}{2 n!} \,.
\end{equation}
Similarly, the determinant in \cref{eq:GeneratingFunctionGaussian} can be expanded as 
\begin{equation}\label{eq:FredholmDeterminant_LogarithmExpansion}
 \det\mleft( \Id + \bm{W} \bm{\varGamma} \mright) = \exp\mleft( -\sum_{n=1}^\infty \frac{(-1)^n}{n} \Tr\mleft[ (\bm{W} \bm{\varGamma})^n \mright] \mright)\,,
\end{equation}
leading to another approximation by truncating the series.

The transformation of the state between photon pair generation and photon detection can be described by an operator $\bm s$ and the selection of a subset of modes to be detected can be described by an orthogonal projection $\bm P$.
By applying Sylvester's determinant theorem~\cite{Pozrikidis_2014}, the expression for the final determinant describing the detection of the photons in \cref{eq:GeneratingFunctionGaussian} can be simplified as
\begin{equation}\label{eq:DeterminantSylvester}
    \det\mleft( \mathds{1}_{2M} + \bm{P} \bm{s} \bm{\varGamma} \bm{s}^\dag \bm{P} \mright) = \det\mleft( \mathds{1}_{2M'} + \bm{s}^\dag \bm{P} \bm{s} \bm{\varGamma} \mright) \,,
\end{equation}
where $M$ is the total number of discrete modes and $M'$ is the number of discrete modes that are not in the vacuum state after the SPDC process.

\section{Setup and Protocol of the QKD Systems}\label{sec:SetupAndTransformations}
We simulate two entanglement-based QKD systems which utilize the time-bin encoded BBM92 protocol~\cite{BBM_1992,Brendel_1999,Tittel_2000} to establish secure symmetric keys for cryptographic applications between pairs of four users Alice, Bob, Charlie and Diana. 
Experimental details about the systems have been presented in refs.~\cite{Fitzke_2022_4Party, Dolejsky_2023} and a simulation without spectral resolution has been presented in \refcite{Fitzke_2023}. 
\begin{figure*}
    \centering
    \includegraphics{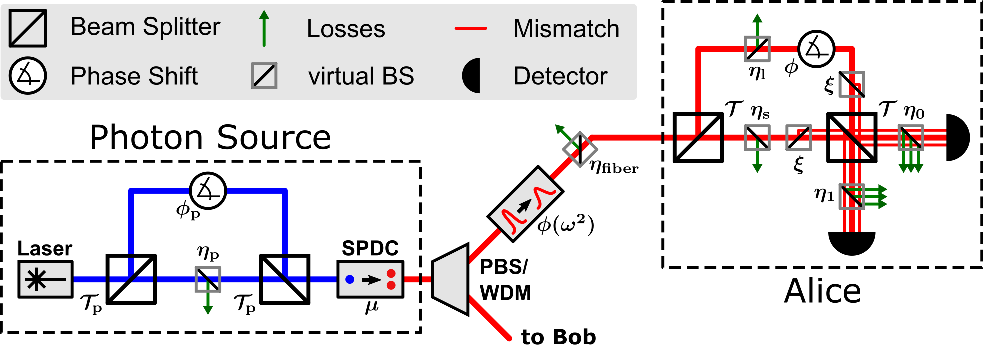}
    \caption{Setup of the QKD system implementing the time-bin encoded BBM92 protocol. A laser emits pulses that are split in two non-overlapping pulses by an imbalanced interferometer with beam splitter transmittivities $\mathcal{T}_\upp$, relative losses $\eta_\upp$ and a phase-shift $\phi_\upp$ between the long and short arm. Both pulses coherently excite an SPDC process such that the mean number of generated photon pairs per (double) pulse is $\mu$. The pairs are split either by a polarizing beam splitter~(PBS) or a wavelength-division demultiplexer~(WDM). The separated photons are sent through optical fibers with a length of several kilometers to the QKD users, for example Alice and Bob. In the fibers they acquire propagation losses $\eta_{\mathrm{fiber}}$ and a quadratic phase $\phi(\omega^2)$ inducing chromatic dispersion. At the receivers, the photons traverse imbalanced interferometers with beam splitter transmittivities $\mathcal{T}$, losses $\eta_\ups, \eta_\upl$, a phase difference $\phi$ and a mode match of $\xi$ w.r.t.\ the pump interferometer. They are detected at both interferometer outputs after experiencing an additional loss of $\eta_0$ or $\eta_1$ due to limited coupling and detection efficiencies. Bob's side of the setup is equivalent to Alice's but features an independent set of parameters.}
    \label{fig:PTC_Setup}
\end{figure*}

The setup is shown in \cref{fig:PTC_Setup}: Nearly transform-limited laser pulses with a length of approximately \SI{0,4}{\nano\second} traverse a pump interferometer, introducing a temporal path-length difference longer than the pulse duration. Thereby, each pulse is split into two separate pulses with a phase difference of $\phi_\upp$ between them. Both pulses excite an SPDC process to produce frequency-entangled photon pairs. The generated pairs are split up such that one photon of each pair travels to one of the users (Alice) and the other one travels to another user (Bob), where they each pass a second interferometer with a path length difference matched to that of the pump interferometer, introducing additional phases $\phi_\upA$ and $\phi_\upB$ between both arms, respectively. Finally, the photons are detected in the outputs of each interferometer. For every pump pulse this leads to three distinct time-bins in which Alice and Bob expect detection events, corresponding to the different combinations of long and short paths in the interferometers. They agree on events in the time basis, i.e.\ events where both of them registered a photon in one of the outer time-bins (early~(e) and late~(l)) and events in the phase basis, i.e.\ coincidences in the central time-bin~(c).\footnote{When the time of flight for the photons is different due to different travel different distances to Alice and Bob, we mean by "coincidence" that the photons are detected with the time difference that is expected when they were generated simultaneously at the source.} All other events are discarded. The key bits in the time basis are obtained from the time bin the respective photon has been detected in ("0" for the early and "1" for the late time bin).
The key bits in the phase basis are obtained from coincident events in the central time-bin, corresponding to the pump photon taking the short path and Alice's and Bob's photons taking the long path in their respective interferometers or vice versa, leading to non-local two-photon Franson interference~\cite{Franson_1989}. In an ideal setting, neglecting the impact of multi-photon-pair generations and all explicitly time/frequency-dependent effects, the probability of observing coincidence events in pairs of detectors $D_\upA$ and $D_\upB$ with $D_\upA, D_\upB \in \{0, 1\}$ of Alice and Bob in the phase basis is proportional to~\cite{Brendel_1999, Tittel_2000}
\begin{equation}\label{eq:ideal_coincidence_probability}
    P_{D_\upA, D_\upB} \propto 1+(-1)^{D_\upA+D_\upB} \cos(\phi_\upA + \phi_\upB - \phi_\upp)\,.
\end{equation}
This means that the phases can be adjusted such that the photons of a pair will always be registered in correlated detectors ($D_\upA = D_\upB$) when both photons are detected in the central time bin, with the respective detector number directly corresponding to the bit value.
Quantum bit errors occur when one photon is detected in the early time bin and the other photon is detected in the late time bin, or when the photons in the central time bin are detected in anti-correlated detectors ($D_\upA \neq D_\upB$).

Prior to the simulations, we thoroughly determined all relevant system parameters by consulting test reports of the optical components, performing measurements of the conversion efficiencies of the SPDC crystals and coupling efficiencies as well as the two-interferometer mode mismatches. The efficiencies, dead times, dark counts and afterpulsing probabilities of the single-photon avalanche detectors were determined by detector tomography~\cite{Fitzke_2022_POVM,Fitzke_2023}. We performed simulations according to a worst- and best-case scenario to take into account the experimental uncertainties of the parameters. The worst-(best-)case scenario uses the lowest (highest) estimates for all transmissions and mode matches as well as the highest (lowest) estimates for the detector dead times and afterpulse probabilities.

Furthermore, effects such as chromatic dispersion or frequency-dependent losses caused by the non-zero spectral linewidths of the photons are relevant and, therefore, included in the simulation. The former typically becomes less important and the latter more important when employing wavelength-division demultiplexing. The spectrum also determines the temporal amplitude of the photon pairs arriving at the participants which impacts the interference visibility as well as the optimal choice of parameters such as the width of the time-bins. Effects from multi-photon-pair events are considered by the covariance formalism for continuous-mode highly entangled states as developed in the first part of the series~\cite{Kleinpass_2024_partI}. Furthermore, we use a model from \refcite{Takeoka_2015} to account for mode mismatches in other degrees of freedom: In each receiver interferometer arm we introduce a virtual beam splitter coupling a part of the intensity into an auxiliary mismatch mode. At the output beam splitter, each mismatch mode is superposed with vacuum and thus does not contribute to the interference. However, here, because of the Franson-type interferometer configuration, the mismatch of the receivers is always defined relative to the pump interferometer.

\section{Simulation Model and Results}\label{sec:Simulation_Model_and_Results}
In section \cref{sec:Photon_Pair_Generation} we present models used to describe the generation of the photon pairs, specifically to obtain the JSAs of the SPDC processes from simple measurements of the pump pulse and the photon pair spectra, as well as, for the \typezero case, the channels of the wavelength-division multiplexer. Afterwards, the transformations describing the transmission of the photons through the optical fibres and receiver interferometers are presented in \cref{sec:Photon_Transmission_through_the_Fiber_Links}, before a detailed model of the single-photon avalanche diodes used for the detection is developed in \cref{sec:Photon_Detection}. Finally, in \cref{sec:Key_Rates} the effect of multi-pair emissions on the key and quantum bit error rates is examined.

\subsection{Photon Pair Generation}\label{sec:Photon_Pair_Generation}

The delay of the pump interferometer is larger than the pulse duration, such that both halves of the pump pulse excite the SPDC process independently. Therefore, the covariance of the biphoton state generated after the pump interferometer becomes
\begin{equation}\label{eq:RenormalizedCovariance_PumpIF}
    \bm{\varGamma} = \bm{\varGamma}_\ups + \bm{\varGamma}_\upl \,.
\end{equation}
Here, $\bm{\varGamma}_\ups$ and $\bm{\varGamma}_\upl$ take the form of \cref{eq:Covariance_Type0} or \cref{eq:Covariance_TypeII}, with the corresponding subnormalized JSAs 
\begin{equation}
    \psi_z(\omega_\ups, \omega_\upi) = K_{\upp,z} \upe^{\upi \phi_{\upp,0}^{(z)}} \upe^{\upi \omega_+ \tau_\upp^{(z)}} \psi(\omega_\ups, \omega_\upi)
\end{equation}
for ${z \in \{\ups, \upl\}}$, where $\phi_{\upp,0}^{(z)}$ and $\tau_\upp^{(z)}$ are the acquired phase shifts and time shifts and the coefficients read ${K_{\upp,\ups} = \mathcal{T}_\upp^2/(\mathcal{T}_\upp^4 + \mathcal{R}_\upp^4)^{1/2}}$ and ${K_{\upp,\upl} = \mathcal{R}_\upp^2/(\mathcal{T}_\upp^4 + \mathcal{R}_\upp^4)^{1/2}}$. The corresponding squeezing parameters are obtained from the normalized JSA $\psi(\omega_\ups, \omega_\upi)$ with squeezing parameters $\bm{\sigma}$ according to ${\bm{\sigma}_z = K_{\upp,z} \bm{\sigma}}$. Both halves of the pulse contribute to the mean number of generated photon pairs~$\mu$.

Neglecting effects from time-ordering~\cite{Quesada_2014}, the JSA of an SPDC process is given by
\begin{equation}
    \psi(\omega_\ups,\omega_\upi) = \alpha(\omega_+) \Phi(\omega_\ups, \omega_\upi) \,,
\end{equation}
where $\alpha(\omega)$ is the amplitude of the coherent pump pulse which, by energy conservation, determines which photon frequencies are allowed. The phase-matching function $\Phi(\omega_\ups, \omega_\upi)$ represents the conservation of momentum. 

For states featuring a large aspect ratio, i.e.\ states where the pump spectrum in ${\omega_+ = \bar{\omega}_\ups + \bar{\omega}_\upi}$ direction is very narrow compared to the variation of the phase-matching function in ${\omega_- = \bar{\omega}_\ups - \bar{\omega}_\upi}$ direction (cf.~\cref{fig:BandJSA_Spread}), the phase-matching function can be approximated as ${\Phi(\omega_\ups, \omega_\upi) \to \Phi(\omega_-)}$ because $\Phi$  is nearly constant over the small interval of $\omega_+$ in which the pump pulse amplitude is non-negligible. Here, ${\bar{\omega}_\ups = \omega_\ups - \omega_{\ups,0}}$ and ${\bar{\omega}_\upi = \omega_\upi - \omega_{\upi,0}}$ are the angular frequencies of signal and idler centered around the respective carrier frequencies $\omega_{\ups,0}$ and $\omega_{\upi,0}$. This approximation leads to a factorization of the JSA
\begin{IEEEeqnarray}{c"c}
    \psi(\omega_\ups, \omega_\upi) \approx \alpha(\omega_+) \Phi(\omega_-)
\end{IEEEeqnarray}
w.r.t. $\omega_+$ and $\omega_-$, corresponding to a rotation of the coordinates by $\SI{45}{\degree}$ in combination with stretching by a factor of $2$. In this approximation, the temporal dynamics are governed by the phase-matching function (in $t_-$ direction) while the pump pulse determines the propagation of the reference time (in $t_+$ direction), similar to center-of-mass and relative motions in classical mechanics. However, effects like chromatic dispersion may introduce non-linear phase factors that do not factorize.

Although the setups using \typeII and \typezero photons are similar, there are some significant differences determining the approach taken for the simulation of the different SPDC processes. 
For the \typeII system, the spectrum of the photons travelling to Alice and Bob is very broad, i.e.\ the JSA features a large aspect ratio and the mean number of photon pairs fulfills ${\mu \ll 1}$. Therefore, the series expansions in \cref{eq:RenormalizedCovariance_SeriesExpansion,eq:FredholmDeterminant_LogarithmExpansion} converge rapidly, allowing us to apply the bivariate Poisson approximation introduced in \refcite{Kleinpass_2024_partI}. As in \typeII processes both photons of a pair end up in different polarization modes, the JSA may exhibit asymmetries. In \cref{sec:TypeII_Asymmetric_JSA}, we introduce a procedure to reconstruct it from independent measurements of the pump and photon pair spectra.

For the \typezero system, the spectrum of the photon pairs is even broader. The wavelength-division demultiplexer, however, selects comparatively narrow frequency channels that are connected to the users. Two users connected to channels that are symmetrically spaced around the center frequency receive entangled photons such that they can generate secret keys. In this way, different wavelength channels may be used to connect multiple pairs of users simultaneously. Therefore, the overall mean number of generated pairs over the whole \typezero spectrum is much larger than for an individual channel pair and the series expansions may be slower to converge. For each pair of users, however, we can restrict the JSA to only those frequency components that effect the state after the multiplexer. This procedure is introduced in \cref{sec:Type0_Wavelength_Division_Demultiplexing}.

\subsubsection{Type-II System: Asymmetric JSA}\label{sec:TypeII_Asymmetric_JSA}
Using the approximate factorization of the JSA in terms of $\alpha(\omega_+)$ and $\Phi(\omega_-)$ allows us to reconstruct the JSA from independent measurements of the pump pulse and photon-pair spectra. Pumping the SPDC process with cw-light of frequency $\omega_0$, corresponding to ${\alpha(\omega_+) = \delta(\omega_0-\omega_+)}$, yields the marginal distributions ${\Psi_\ups(\omega_\ups) \propto \abs{\Phi(2\omega_\ups-\omega_0)}^2}$ and ${\Psi_\upi(\omega_\upi) \propto \abs{\Phi(\omega_0-2\omega_\upi)}^2}$ for signal and idler, respectively. Thus, measuring the frequency-resolved count rate of the photons in the low gain regime (${C \ll 1}$ in \cref{eq:SPDC_linear}) provides $\abs{\Phi(\omega_-)}$.

\begin{figure*}
    \centering
    \includegraphics{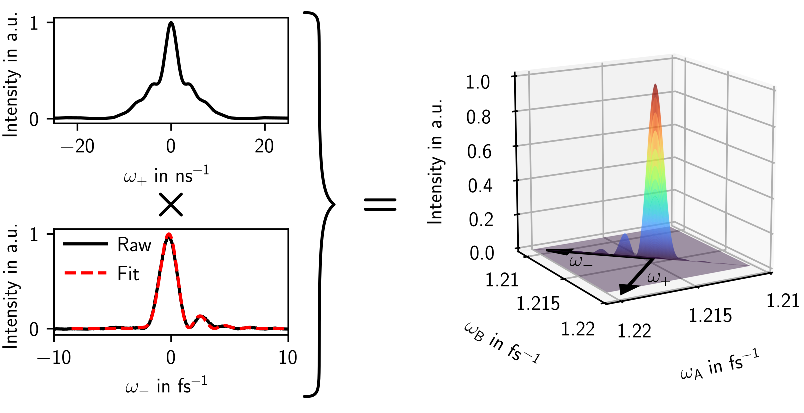}
    \caption{JSD reconstructed from measurements of the pump pulse spectrum in $\omega_+$-direction and the \typeII photon-pair spectra in $\omega_-$-direction. To account for the asymmetry in $\omega_-$-direction, \cref{eq:type2_spectrum_model} was fitted to the pair spectra.}
    \label{fig:TypeII_JSD}
\end{figure*}

In experimental realizations, the spectrum of the generated photon pairs can differ significantly from the expected $\sinc^2$-shape predicted by simple theoretical models~\cite{Ou_2007_Multiphoton} due to imperfections during the manufacturing process of the non-linear crystal. For example, the measured \typeII spectrum in our QKD system exhibits a significant asymmetry of the side lobes as shown in~\cref{fig:TypeII_JSD}. Details about the measurements of the photon spectra of both the \typeII and \typezero crystals have been reported in \refcite{Euler_2021}.

In addition to the absolute value $\abs{\Phi(\omega_-)}$ obtained by taking the square root of the measured spectrum, the correct frequency-dependent phase of $\Phi(\omega_-)$ is required. Therefore, we reconstruct $\Phi(\omega_-)$ from the model described below. The asymmetry of the measured \typeII spectrum is most likely due to imperfections of the non-linear crystal waveguide, for example slight variations of the effective refractive indices leading to a spatial dependence of the phase-mismatch $\Updelta k(\omega_-, z)$ along the waveguide direction $z$~\cite{Helmfrid_91,Gray_2020}. Following  \refcite{Chang_2014,Santandrea_2019}, we write the phase-matching function $\Phi(\omega_-)$ for a crystal of length $L$ as 
\begin{equation}\label{eq:Phase_matching_ansatz}
    \Phi(\omega_-) \propto \int_{-L/2}^{L/2} \exp\mleft(\upi \int_{-L/2}^z \Updelta k(\omega_-, \xi) \dl \xi\mright) \dl z\,.
\end{equation}
By using the decoupling approximation~\cite{Santandrea_2019}
\begin{equation}\label{eq:decoupling_approximation}
    \Updelta k(\omega_-, z) \approx \Updelta k(\omega_-) + \delta k(z) \,,
\end{equation} 
the phase mismatch is split into a frequency-dependent contribution $\Updelta k(\omega_-)$, which is uniform along the crystal, and a small perturbation $\delta k(z)$ accounting for spatial waveguide imperfections affecting all frequencies equally~\cite{Chang_2014,Santandrea_2019}.

By inserting \cref{eq:decoupling_approximation} into \cref{eq:Phase_matching_ansatz}, the phase-matching function can be written as the Fourier transform of a position-dependent phase factor and a rectangle function representing the crystal of length~$L$~\cite{Santandrea_2019}:
\begin{equation}\label{eq:Phasematching_FourierTransform}
    \Phi(\omega_-) \propto \mathcal{F}_z\mleft[\exp\mleft( \upi \int_{-L/2}^z \delta k(\xi) \dl \xi \mright) \mathrm{rect}_L(z)\mright]\mleft[\Updelta k(\omega_-)\mright] \,.
\end{equation}
For a uniform crystal, ${\delta k(\xi) = 0}$ and the phase-matching function takes the form ${\Phi(\omega_-) \propto \sinc[\Updelta k(\omega_-) L / 2]}$.

Assuming that the pertubation varies smoothly along the crystal, we expand it up to the second order around the crystal center as
\begin{equation}
    \delta k(\xi) = \delta k' \xi +  \frac{1}{2}\delta k'' \xi^2 \,,
\end{equation} 
where the position-independent term has been absorbed into $\Updelta k(\omega_-)$. When the SPDC is pumped with continuous-wave~(cw) laser light, this yields the spectral density 
\begin{equation}\label{eq:type2_spectrum_model}
    |\Phi(\omega_-)|^2 \propto \abs*{\mathcal F_z\mleft[\upe^{\mathrm i (\delta k' z^2/2 + \delta k'' z^3/6)} \rect_{L}(z) \mright]\mleft[\Updelta k(\omega_-)\mright]}^2 \,.
\end{equation}
The unknown coefficients ${\delta k'}$ and ${\delta k''}$ describe the lowest-order symmetric and antisymmetric corrections to the ideal $\sinc\textsuperscript2$-shape, respectively.

For the reconstruction of the phase-matching function, \cref{eq:type2_spectrum_model} can be fitted to the measurement data to obtain ${\delta k'}$ and ${\delta k''}$ as shown in \cref{fig:TypeII_JSD}. The phase matching function is then obtained from \cref{eq:Phasematching_FourierTransform}.

In practice, to obtain an expression for ${\Updelta k(\omega_-)}$, we perform another expansion up to the linear term ${\Updelta k(\omega_-) \approx \Updelta k_0 + \Updelta k' \omega_-}$. The width of the ideal $\sinc$ peak is determined by the parameter ${\Updelta k'}$, which we calculate from the Sellmeier equations for LiNbO\textsubscript 3~\cite{Gayer_2008} and the crystal length of $L = \SI{24}{\milli\meter}$. The position of the peak, determined by ${\Updelta k_0}$, is governed by different effects, such as the quasi-phase-matching poling period, the crystal temperature and the properties of the wave guides. Therefore, we also fit the value of ${\Updelta k_0}$ as well as a constant background to adjust the baseline of the measured spectrum to zero. 

\subsubsection{Type-0 System: Wavelength-Division Demultiplexing}\label{sec:Type0_Wavelength_Division_Demultiplexing}

\begin{figure}
    \centering
    \includegraphics{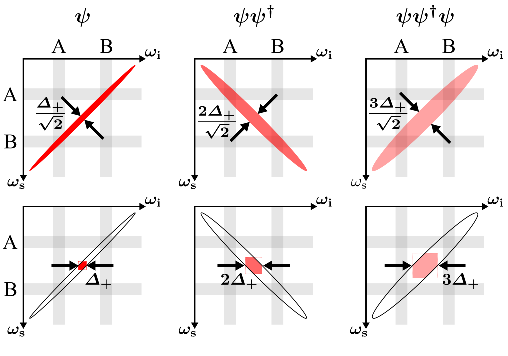}
    \caption{Visualization of the kernels of the operators iterated in $\bm{\psi}$, which attain non-negligible values only on a narrow stripe, schematically indicated by the oval areas, with large aspect ratio. For higher orders, more frequency components become relevant. However, for expansion order $\mathcal{O}(\psi^n)$, only the frequency components in a vicinity of ${n \varDelta_+ /2}$ (red areas in lower row) affect the $n$-th iterated kernel.}
    \label{fig:BandJSA_Spread}
\end{figure}
\begin{figure}
    \centering
    \includegraphics{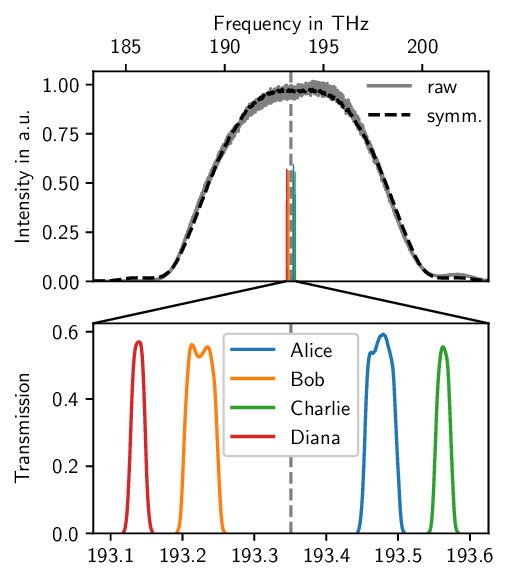}
    \caption{Measurements of the \typezero SPDC spectrum (top) and the transmission functions of the wavelength-selective switch used as WDM (bottom). As signal and idler are generated in the same polarization mode, they are indistinguishable and the spectrum needs to be symmetric. The asymmetry of the measured spectrum is due to wavelength dependencies of the spectrograph. For the simulation the measured spectrum was symmetrized (dashed line in upper plot). The nominal WDM channel widths are $\SI{50}{\giga\hertz}$ for Alice and Bob and $\SI{25}{\giga\hertz}$ for Charlie and Diana.}
    \label{fig:Type-0_WSS}
\end{figure}
\begin{figure}
    \centering
    \includegraphics{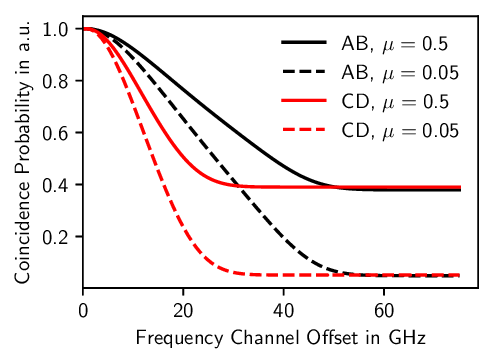}
    \caption{Simulated normalized coincidence detection probability between Alice and Bob (AB) as well as Charlie and Diana (CD) corresponding to the frequency channels from \cref{fig:Type-0_WSS} when introducing an additional offset of the central frequency of Alice' and Charlie's channel.}
    \label{fig:CoincidenceProb_ChannelAsymmetry}
\end{figure}
For many applications, it is beneficial or even necessary to split signal and idler photons according to their wavelength rather than their polarization. This is most conveniently realized by utilizing highly entangled states generated by a \typezeroI process with a sufficiently large aspect ratio of the JSA. The narrow pump spectrum leads to signal and idler photons being generated symmetrically around a sufficiently sharp center frequency, while the broad phase-matching function allows to distribute photons to multiple channel pairs simultaneously and allows for the wavelength channels to be sufficiently far away from each other to prevent both photons of a pair from entering the same channel. Note that if the frequency channels are much more narrow than the width of the SPDC spectrum, the total number of photon pairs over the whole frequency range might be much larger than the number of photons per channel.

The large aspect ratio of the JSA leads to the same computational difficulties as described above, while in addition in this case the total mean photon number may be too large for the Poisson approximation (cf.~\refcite{Kleinpass_2024_partI}) to be applied. Higher-order expansions require the evaluation of higher-order iterations of $\bm{\psi}$ or an explicit Schmidt decomposition, both of which become numerically challenging due to the narrow structure of the JSA in $\omega_+$ direction over the broad spread in $\omega_-$ direction. Ultimately, however, only the state within the wavelength channels is of interest. As it will be discussed below, for highly entangled states only the frequency components in a close vicinity of the wavelength channels are relevant. Thus, the JSA can be restricted to the interval of contributing frequency components where the methods discussed earlier may be applied, i.e.\ the Schmidt decomposition or higher orders of the reduced JSA may be computed explicitly. A similar method was already discussed in~\refcite{Nauth_2022}, however not in the context of wavelength-division demultiplexing.

Consider a state with a full width of $\varDelta_+$ in the $\omega_+$ direction, that is, a state with a JSA $\psi(\omega_\ups, \omega_\upi)$ that is only non-zero if ${\abs{\omega_+} \leq \varDelta_+/2}$. The $n$-th order contribution to the series expansion of its covariance in \cref{eq:RenormalizedCovariance_SeriesExpansion} is governed by terms of the form
\begin{equation}\label{eq:JSA_n_Definition}
    \psi_n(\omega_1, \omega_{n+1}) = \int \dl \omega_2 \dots \dl \omega_n \psi(\omega_1, \omega_2) \psi^*(\omega_3, \omega_2) \dots \,.
\end{equation}
Due to the width in $\omega_+$ direction, the product under the integral can only be non-zero where ${\abs{\bar{\omega}_j + \bar{\omega}_{j+1}} \leq \varDelta_+/2}$ for all ${j = 1, \dots, n}$. Therefore, $\psi_n(\omega_1, \omega_{n+1})$ can only be non-zero if
\begin{equation}
    \abs{\bar{\omega}_1 - (-1)^n \bar{\omega}_{n+1}} \leq n \varDelta_+/2 \,, 
\end{equation}
meaning that with each order~$n$ of the covariance expansion, the corresponding contribution $\psi_n(\omega_1, \omega_{n+1})$ features a full width of ${n \varDelta_+}$ w.r.t. $\omega_+$ or $\omega_-$, alternating between an anti-diagonal and a diagonal shape for odd and even values of $n$ in the coordinates $\omega_\ups$ and $\omega_\upi$, respectively, as shown in \cref{fig:BandJSA_Spread}.

The frequency transmission channels of two receivers Alice $(\upA)$ and Bob $(\upB)$ are assumed to be non-zero only within the intervals ${[\overline{a}, \underline{a}]}$ and ${[\underline{b}, \overline{b}]}$, respectively, fulfilling ${\overline{a} \leq \underline{a} \leq 0 \leq \underline{b} \leq \overline{b}}$. After the wavelength-division demultiplexing, $\psi_n(\omega_1, \omega_{n+1})$ can only be non-zero if ${(\bar{\omega}_1, \bar{\omega}_{n+1})}$ is contained within one of the four rectangular intervals
\begin{subequations}\label{eq:FrequencyChannelsWDM}
    \begin{IEEEeqnarray}{cCc"cCc}
        (\upA\upA)&:& [\overline{a}, \underline{a}] \times [\overline{a}, \underline{a}] \,, &
        (\upA\upB)&:& [\overline{a}, \underline{a}] \times [\underline{b}, \overline{b}] \,,
        \IEEEeqnarraynumspace\\
        (\upB\upA)&:& [\underline{b}, \overline{b}] \times [\overline{a}, \underline{a}] \,, & 
        (\upB\upB)&:& [\underline{b}, \overline{b}] \times [\underline{b}, \overline{b}] \,,
        \IEEEeqnarraynumspace
    \end{IEEEeqnarray}
\end{subequations}
corresponding to the intersection (darker gray squares) of the wavelength channels (light gray rectangles) in \cref{fig:BandJSA_Spread}.

The anti-diagonal elements of the covariance are generated by two-mode squeezing operations and describe the presence of correlated photon pairs. They are composed of terms with odd values of $n$ in \cref{eq:JSA_n_Definition} and thus feature an anti-diagonal shape. Therefore, they mainly contribute to intensity in the intersection of Alice's and Bob's channels (see \cref{fig:BandJSA_Spread}), corresponding to the conditions $(\upA\upB), (\upB\upA)$ in \cref{eq:FrequencyChannelsWDM}. These terms represent contributions where signal and idler are split between the channels of both receivers. If the width $n \varDelta_+$ becomes too large, however, the anti-diagonal elements can also reach the intervals $(\upA\upA)$ and  $(\upB\upB)$, leading to a non-zero probability of signal and idler both ending up in the same wavelength channel.

Denoting the innermost and outermost channel bound as ${\underline{c} = \min(-\underline{a}, \underline{b})}$ and ${\overline{c} = \max(-\overline{a}, \overline{b})}$, respectively, the condition
\begin{equation}\label{eq:WDM_Condition_NoDoublePhotons}
    \underline{c} > N \varDelta_+ / 4
\end{equation}
is sufficient to prevent both photons of a pair from entering the same wavelength channel. Here, $N$ is the highest contributing order to the series expansion of the covariance in \cref{eq:RenormalizedCovariance_SeriesExpansion}. Furthermore, the conditions on the width of the JSA in \cref{eq:JSA_n_Definition} can be used to show that only frequency components fulfilling
\begin{IEEEeqnarray}{c}\label{eq:FrequencyCondition_WDM}
    \underline{c} - N \varDelta_+ / 2 \leq \abs{\bar{\omega}_j} \leq \overline{c} + N \varDelta_+ / 2
\end{IEEEeqnarray}
can contribute to the intensity within the wavelength channels. For symmetrically chosen channels, this means that only frequencies in a vicinity of ${N\varDelta_+/2}$ around the channel bounds can have an impact on the $N$-th order kernel. Thus, the covariance describing the state within the two wavelength channels of Alice and Bob is completely determined by the reduced JSA
\begin{equation}
    \overline{\bm{\psi}} = \bm{\mathcal{P}} \bm{\psi} \bm{\mathcal{P}} \,,
\end{equation}
where $\bm{\mathcal{P}}$ is the orthogonal projection onto the frequencies fulfilling \cref{eq:FrequencyCondition_WDM}. For comparatively narrow wavelength channels, the reduced JSA $\overline{\bm{\psi}}$ features a much smaller aspect ratio and requires fewer discretization points to be resolved to sufficient accuracy, making it much easier to compute higher orders of the covariance expansion. The renormalized covariance after the wavelength-division can be written as
\begin{equation}
    \bm{\varGamma}' = \bm{\eta} \bm{\varGamma} \bm{\eta} = \bm{\eta} \bm{P} \bm{\varGamma} \bm{P} \bm{\eta} = \sum_{n=1}^N \frac{(2C)^n}{2n!} \bm{\eta} \mqty{ 0 & \overline{\bm{\psi}}\\ \overline{\bm{\psi}}^\dag & 0 }^n \bm{\eta} \,,
\end{equation}
where ${\bm{P} = \bm{\mathcal{P}}^{\oplus 2}}$ and
\begin{equation}
    \bm{\eta} = \mqty{ T_\upA(\omega) & 0\\ 0 & T_\upB(\omega) }^{\oplus2}
\end{equation}
represents the transmission through the wavelength channels of the receivers Alice and Bob, modeled as frequency-dependent loss transformations with the transmission functions $T_{\upA}^2(\omega)$ and $T_\upB^2(\omega)$, respectively. Equivalently, the renormalized covariance may be written in terms of the Schmidt decomposition of the reduced JSA ${\overline{\bm{\psi}} = \bm{U} \bm{\Sigma} \bm{V}^\dag}$, which yields
\begin{equation}
      \bm{\varGamma}' = \frac{\bm{\eta}}{2}
    \mqty{ 
       \bm{U} \mleft[ \mathfrak{c}_N(\bm{\sigma}^{(\upI)}) - \mathds{1} \mright] \bm{U}^\dag &
        \bm{U} \mathfrak{s}_N(\bm{\sigma}^{(\upI)}) \bm{V}^\dag\\
        \bm{V} \mathfrak{s}_N(\bm{\sigma}^{(\upI)}) \bm{U}^\dag &
        \bm{V} \mleft[ \mathfrak{c}_N(\bm{\sigma}^{(\upI)}) - \mathds{1} \mright] \bm{V}^\dag
    }
    \bm{\eta} \,.
\end{equation}
If the channels are chosen such that \cref{eq:WDM_Condition_NoDoublePhotons} is fulfilled, the probability of both photons of a pair entering the same channel vanishes and the covariance can be reduced further by reordering the basis elements (cf.~\refcite{Kleinpass_2024_partI}), i.e.\ ${\bm{\varGamma}' \to \bar{\bm{\varGamma}} \oplus \cc}$ with
\begin{equation}\label{eq:Covariance_Type0_WDM}
    \bar{\bm{\varGamma}} = \frac{1}{2}
    \mqty{
        \bm{U}_\upA \mleft[ \mathfrak{c}_N(\bm{\sigma^{(\upI)}}) - \mathds{1} \mright] \bm{U}_\upA^\dag &
        \bm{U}_\upA \mathfrak{s}_N(\bm{\sigma^{(\upI)}}) \bm{V}_\upB^\dag\\
        \bm{V}_\upB \mathfrak{s}_N(\bm{\sigma^{(\upI)}}) \bm{U}_\upA^\dag &
        \bm{V}_\upB \mleft[ \mathfrak{c}_N(\bm{\sigma^{(\upI)}}) - \mathds{1} \mright] \bm{V}_\upB^\dag
    } \,,
\end{equation}
where
\begin{IEEEeqnarray}{c"c}
    \bm{U}_\upA = \bm{T}_\upA \bm{U} \,, &
    \bm{V}_\upB = \bm{T}_\upB \bm{V} \,.
    \IEEEeqnarraynumspace
\end{IEEEeqnarray}
Here, $\bm{U}_\upA$ and $\bm{V}_\upB$ are in general not unitary due to the multiplication by the transmittivity functions.

Applying those considerations to our setup, we truncate the series expansion in \cref{eq:RenormalizedCovariance_SeriesExpansion} at order ${N = 5}$, limiting the JSA to a vicinity of ${5 \varDelta_+/2}$ around the wavelength channels. On this interval, we can perform a discretization using a reasonably small amount of grid points. In this case, the matrix dimension is sufficiently low to directly evaluate the determinant of the matrix numerically.

\Cref{fig:Type-0_WSS} shows measurements of the \typezero SPDC spectrum as well as the channels of the multiplexer employed in our QKD setup described in \cref{sec:SetupAndTransformations}. The asymmetry in the measurement is due to the frequency-dependent detection efficiency of the spectrograph. Because of energy conservation, however, the actual spectrum of the generated photons must be symmetric. Therefore, the measured spectrum is symmetrized for the simulation.

In \cref{fig:CoincidenceProb_ChannelAsymmetry} the impact of multi-pair events and asymmetric frequency channels on coincidence detection rates is examined. If both channels are chosen symmetrically spaced around half the central pump frequency the coincidence probability takes on its maximum. When shifting one of the channels w.r.t.\ the center frequency the correlations decrease until the coincidence probability approaches the product of the independent detection probabilities. For larger values of $\mu$, the increased amount of multi-photon-pair emissions leads to additional uncorrelated detections, decreasing the overall contrast between correlated and uncorrelated events even for symmetrically spaced frequency channels.

\subsection{Photon Transmission through the Fiber Links}\label{sec:Photon_Transmission_through_the_Fiber_Links}
The transformation describing the setup after the photon-pair generation reads
\begin{equation}\label{eq:RenormalizedCovariance_TransformationBBM92}
    \bm{\varGamma} \to \bm{S}_0 \bm{S}_\omega \bm{\varGamma} \bm{S}_\omega^\dag \bm{S}_0^\dag \,,
\end{equation}
where the total transformation was split into ${\bm{S}_0 = \bm{\eta}_{\mathrm{D}} \bm{\BS}_{\mathrm{IF}}^\dag \bm{\BS}_{\mathrm{MM}} \bm{\eta}_{\mathrm{IF}} \bm{\PS}_{\mathrm{IF}} \bm{\BS}_{\mathrm{IF}} \bm{\eta}_{\mathrm{fiber}}}$, containing only frequency-independent terms and linear phases, and ${\bm{S}_\omega = \bm{\PS}_{\mathrm{CD}} \bm{\eta}_{\mathrm{WDM}}}$, containing the frequency-dependent losses $\bm{\eta}_{\mathrm{WDM}}$ caused by the wavelength-division channels, and the quadratic phase $\bm{\PS}_{\mathrm{CD}}$ due to chromatic dispersion. The general structure of the transformations representing beam splitters $\bm{\BS}$, phase rotations $\bm{\PS}$ and losses $\bm{\eta}$ is described in \refcite{Kleinpass_2024_partI}. 
The transformation $\bm{\eta}_{\mathrm{fiber}}$ in $\bm{S}_0$ represents the loss experienced within the optical fibers. For simplicity, a wavelength-independent loss transmittivity ${\eta_{\mathrm{fiber}} = \exp[-L/(2L_0)]}$ with the characteristic loss distance ${L_0 = 10^3/\ln(10^{\alpha \cdot \SI{0.1}{\kilo\meter}})}$ is assumed, where ${\alpha \approx \SI[per-mode=symbol]{0.2}{\deci\bel\per\kilo\meter}}$ is the loss coefficient. A wavelength-dependency could be absorbed into the transmission function of the WDM if required. The transformations $\bm{\BS}_{\mathrm{IF}}, \bm{\PS}_{\mathrm{IF}}$, and $\bm{\eta}_{\mathrm{IF}}$ describe the beam splitters, phase-shifts, and losses within the receiver interferometers. The transformation $\bm{\BS}_{\mathrm{MM}}$ models the virtual mode mismatch beam splitters. Another loss transformation $\bm{\eta}_{\mathrm{D}}$ accounts for additional coupling losses and non-unity detection efficiencies. Due to the Michelson geometry of the interferometers employed in the setup, the output beam splitter is described by the Hermitian adjoint of the input beam splitter.

After the photon-pair generation, we only have two discrete degrees of freedom (DOFs), corresponding to the photons in Alice's and Bob's fibers.\nolinebreak\footnote{The polarization DOF is not modeled explicitly as the polarization mode dispersion is negligible and the interferometers include a Faraday mirror compensating polarization mismatches between the interferometer paths (see ref.~\onlinecite{Fitzke_2022_4Party}). Modeling the  polarization would double the number of discrete DOFs.} Modeling the interferometers including the mode mismatch model introduces in total 10 additional discrete DOFs which are initially in the vacuum state. The expressions can be significantly simplified by applying \cref{eq:DeterminantSylvester} as demonstrated in \cref{sec:Appendix_Analytical_Expressions_for_the_Detection_Probabilities}, yielding the reduced transformation
${\bar{\bm{s}} = \bar{\bm{s}}_\upA \oplus \bar{\bm{s}}_\upB^*}$, where
\begin{equation}\label{eq:ReducedTransformation_BBM92}
    \bar{\bm{s}}_\rho =
    \mqty{
        \upe^{\upi \phi_\rho^{(\ups)}(\omega_\rho)} \eta_{\rho,0}^{(\ups)} \xi_\rho \mathcal{T}_\rho^2 + \upe^{\upi \phi_\rho^{(\upl)}(\omega_\rho)} \eta_{\rho,0}^{(\upl)} \xi_\rho \mathcal{R}_\rho^2\\
        \upe^{\upi \phi_\rho^{(\ups)}(\omega_\rho)} \eta_{\rho,1}^{(\ups)} \xi_\rho \mathcal{T}_\rho \mathcal{R}_\rho - \upe^{\upi \phi_\rho^{(\upl)}(\omega_\rho)} \eta_{\rho,1}^{(\upl)} \xi_\rho \mathcal{T}_\rho \mathcal{R}_\rho\\
        \upe^{\upi \phi_\rho^{(\ups)}(\omega_\rho)} \eta_{\rho,0}^{(\ups)} \bar{\xi}_\rho \mathcal{T}_\rho^2\\
        \upe^{\upi \phi_\rho^{(\upl)}(\omega_\rho)} \eta_{\rho,0}^{(\upl)} \bar{\xi}_\rho \mathcal{R}_\rho^2\\
        \upe^{\upi \phi_\rho^{(\ups)}(\omega_\rho)} \eta_{\rho,1}^{(\ups)} \bar{\xi}_\rho \mathcal{T}_\rho \mathcal{R}_\rho\\
        -\upe^{\upi \phi_\rho^{(\upl)}(\omega_\rho)} \eta_{\rho,1}^{(\upl)} \bar{\xi}_\rho \mathcal{T}_\rho \mathcal{R}_\rho
    }
\end{equation}
for ${\rho \in \{\upA, \upB\}}$. Here, $\eta_{\rho,D}^{(x)}$ is the total frequency-independent transmittivity a photon experiences when being detected in detector $D$ after taking path ${x \in \{ \ups, \upl \}}$ in the interferometer of party $\rho$. The transmittivity and reflectivity of the virtual mode mismatch and real interferometer beam splitters are given by $\xi_\rho$ and ${\bar{\xi}_\rho = (1-\xi_\rho^2)^{1/2}}$, as well as $\mathcal{T}_\rho$ and ${\mathcal{R}_\rho = (1-\mathcal{T}_\rho^2)^{1/2}}$, respectively.

The reduced transformation $\bar{\bm{s}}$ is of shape $6 \times 1$ and effectively reduces the expression to two discrete modes again, in the sense of the operator being evaluated only containing two discrete modes, one corresponding to Alice and the other one to Bob. Each mode contains a sum over the contributions of the different detectors and time intervals determined by the orthogonal projections. An analytic expression of the reduced form of the final renormalized covariance after applying all transformations is given in \cref{eq:FinalCovariance_BBM92}.

\subsection{Photon Detection}\label{sec:Photon_Detection}
The photon detection in the different time bins is modeled by POVMs, taking into account detector imperfections such as dark counts, dead times and afterpulses. The POVM element of detector $D$ associated to a vacuum detection over time interval $I$ is given by
\begin{equation}\label{eq:vac_POVM}
    \hat{\Uppi}^{(D)}_{I,\vac} = \upe^{-r_{\mathrm{noise}}^{(D)} \abs{I}} \outerp{0}{0}_I \,,
\end{equation}
where $\outerp{0}{0}_I$ is the vacuum state over the interval $I$ of length $\abs{I}$ and $r_{\mathrm{noise}}^{(D)}$ is the noise detection rate. The noise is modeled as a Poissonian process~\cite{Barnett_1998}.
The detection probabilites are then obtained as expectation valus of the POVM elements by using \cref{eq:GeneratingFunctionGaussian,eq:DeterminantSylvester,eq:vac_POVM}, yielding 
\begin{equation}
    \big\langle \hat{\Uppi}_{I,\vac}^{(D)} \big\rangle =  \frac{\upe^{-r_{\mathrm{noise}}^{(D)} \abs{I}}}{\abs{\det(\mathds{1} + \tilde{\bm{s}}^\dag \bm{P}_I^{(D)} \tilde{\bm{s}} \tilde{\bm{\varGamma}})}}\,,
\end{equation}
with the time-domain projection operators $\bm{P}_I^{(D)}$ selecting the relevant modes for the time interval $I$ and detector $D$ and the reduced renormalized covariance in the time-domain $\tilde{\bm{\varGamma}}$ as defined in \cref{eq:ReducedTransform_TimeCovariance}.

The total rate of uncorrelated noise detections $r_{\mathrm{noise}}^{(D)}$ consists of dark counts and afterpulses in the single photon avalanche diodes (SPADs):\nolinebreak\footnote{Afterpulses are correlated to prior detection events. However, the afterpulses are distributed over a time interval which is 2 to 3 orders of magnitude longer than the time bin duration (see~\refcite{Fitzke_2022_POVM}), which dilutes the correlations.}
\begin{equation}\label{eq:NoiseRate}
    r_{\mathrm{noise}}^{(D)} = r_{\mathrm{dc}}^{(D)} + r_{\mathrm{ap}}^{(D)} \,.
\end{equation}
The afterpulse rate is given by ${r_{\mathrm{ap}}^{(D)} = p_{\mathrm{ap}}^{(D)} r_{\mathrm{event}}^{(D)}}$, where $p_{\mathrm{ap}}^{(D)}$ is the total probability that, given a detection event, an afterpulse occurs and ${r_{\mathrm{event}}^{(D)} = r_{\mathrm{noise}}^{(D)} + r_{\mathrm{ph}}^{(D)}}$ is the total detection rate comprising noise and events from detected photons $r_{\mathrm{ph}}^{(D)}$. The corresponding photon detection rate reads ${r_{\mathrm{ph}}^{(D)} = r_\upp \langle \hat{n}^{(D)} \rangle}$, where ${\langle \hat{n}^{(D)} \rangle = \Tr\big(\bm{\varGamma}_{\mathrm{final}}^{(D)}\big)/2}$ is the mean number of photons arriving at detector $D$ for each repetition, given by the trace over the corresponding modes, and $r_\upp$ is the pump pulse repetition rate. Solving \cref{eq:NoiseRate} for the noise rate yields
\begin{equation}
    r_{\mathrm{noise}}^{(D)} = \frac{r_{\mathrm{dc}}^{(D)} + p_{\mathrm{ap}}^{(D)} r_\upp \langle \hat{n}^{(D)} \rangle}{1-p_{\mathrm{ap}}^{(D)}} \,.
\end{equation}

Following each detection event the corresponding single-photon detector enters a dead time of ${\tau_{\mathrm{dt}}^{(D)} \sim \SI{10}{\micro\second}}$ during which it is insensitive, before being reactivated. At least one detector at each receiver needs to be active for a coincidence event between Alice and Bob to be detected. Statistically, the probability of a given detector being active at some point in time can be modeled as \cite{Leo_1994}
\begin{equation}
    P_{\mathrm{live}}^{(D)} = \frac{1}{1 + r_{\mathrm{click}}^{(D)} \tau_{\mathrm{dt}}^{(D)}} \,.
\end{equation}
Due to the time-bin encoding, we need to obtain the detection probabilities over the three disjoint time intervals of interest. In general, time intervals between the time-bins may remain unused, meaning that the pulse cycle time is longer than the sum of the durations of the three time-bins.

The security of the key distribution benefits from considering only those events that are registered when all four detectors are active. To ensure that all detectors have been live for a given repetition, the experimental realization post-selects repetitions where, at the beginning of the early time-bins, none of the four detectors is in the dead time.

The dead times of the detectors are longer than the sum of the temporal interferometer mismatches, such that detections in the central time interval $I_\upc$ can never be observed if there was a detection in the early time interval $I_\upe$ and detections in the late time interval $I_\upl$ can never be observed if there was a detection in ${I_\upe \cup I_\upc}$. Thus, the POVM elements describing the actual detection events within the corresponding intervals read
\begin{equation}
    \hat{\Uppi}_{I}^{(D)} = \hat{\Uppi}_{\underline{I},\vac}^{(D)} \mleft( \Id - \hat{\Uppi}_{I,\vac}^{(D)} \mright) \,,
\end{equation}
where $\underline{I_\upe} = \emptyset, \underline{I_\upc} = I_\upe$ and $\underline{I_\upl} = I_\upe \cup I_\upc$.

A photon pair detection that can contribute to the key generation consists of a coincidence click between two detectors $D_\upA, D_\upB$ of Alice and Bob in the time intervals ${(I_\upA, I_\upB) \in \{ (I_\upe, I_\upe), (I_\upe, I_\upl), (I_\upc, I_\upc), (I_\upl, I_\upe), (I_\upl, I_\upl) \}}$ and the simultaneous detection of vacuum in the other two detectors $\bar{D}_\upA, \bar{D}_\upB$ over the whole time interval ${\underline{I} = I_\upe \cup I_\upc \cup I_\upl}$. Thus, the POVM element describing a successful raw key generation event reads
\begin{equation}
    \hat{\Uppi}_{\mathrm{key}}^{(D_\upA,D_\upB)} = \hat{\Uppi}_{\underline{I},\vac}^{(\bar{D}_\upA)} \hat{\Uppi}_{\underline{I},\vac}^{(\bar{D}_\upB)} \hat{\Uppi}_{I_\upA}^{(D_\upA)} \hat{\Uppi}_{I_\upB}^{(D_\upB)} = \hat{\Uppi}_{\underline{I},\vac}^{(\bar{D}_\upA)} \hat{\Uppi}_{\underline{I},\vac}^{(\bar{D}_\upB)} \bigg[ \mathds{1} - \hat{\Uppi}_{\underline{I_\upA} \cup I_\upA,\vac}^{(D_\upA)} - \hat{\Uppi}_{\underline{I_\upB} \cup I_\upB,\vac}^{(D_\upB)} + \hat{\Uppi}_{\underline{I_\upA} \cup I_\upA,\vac}^{(D_\upA)} \hat{\Uppi}_{\underline{I_\upB} \cup I_\upB,\vac}^{(D_\upB)} \bigg]\,.
\end{equation}
The corresponding detection probabilities including the statistical dead time corrections are given by
\begin{equation}
    P_{\mathrm{key}}^{(D_\upA,D_\upB)} = P_{\mathrm{live}}^{(D_\upA)} P_{\mathrm{live}}^{(D_\upB)} P_{\mathrm{live}}^{(\overline{D}_\upA)} P_{\mathrm{live}}^{(\overline{D}_\upB)} \langle \hat{\Uppi}_{\mathrm{key}}^{(D_\upA,D_\upB)} \rangle \,.
\end{equation}
The projections contained in the POVM elements directly translate to the orthogonal projections over the corresponding intervals, acting on the renormalized covariance.

\begin{figure}
    \centering
    \includegraphics{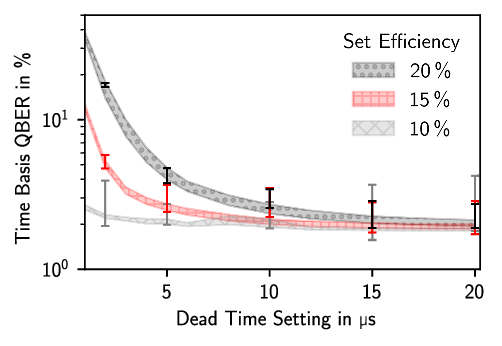}
    \caption{Measured (error bars) and simulated (error bands) Type-0 time basis QBER over different dead time settings for ${\mu = 0.036}$, ${r_\upp = \SI{220}{\mega\hertz}}$ and $\SI{50}{\giga\hertz}$ wide channels (cf.~\cref{fig:Type-0_WSS}).}
    \label{fig:QBER_DeadTime_Type0}
\end{figure}

The single-photon detectors of the QKD system have adjustable dead times (\SIrange[range-units=single]{1}{25}{\micro\second}) and efficiencies (\SIlist[list-units=single]{10;15;20}{\percent}). Higher efficiencies and shorter dead times lead to higher detection rates but also to higher noise levels due to increased afterpulse probabilities and dark count rates. Increasing the dead time reduces this noise but lowers the detection rate, as the detector may be inactive when the next photon arrives.
To investigate the dependency of the noise level on the detector settings we simulated the QBER as a function of the different efficiency and dead time settings using measured dark count rates and afterpulse probabilities. 
\Cref{fig:QBER_DeadTime_Type0} shows a comparison between simulated and measured values of the time basis QBER. The simulation results are in agreement with the measured data. 

\begin{figure}
    \centering
    \includegraphics{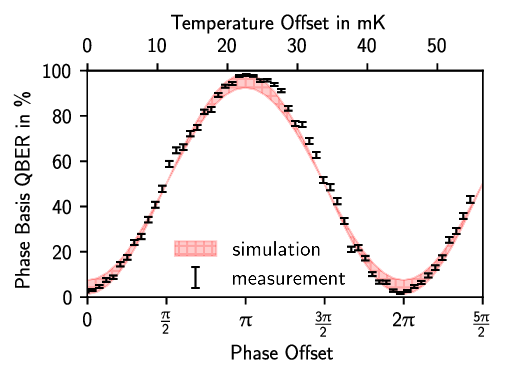}
    \caption{\TypeII phase basis QBER when changing the temperature (i.e.\ phase) of Bob's interferometer. A cosine fit was applied to convert the temperature changes applied to the interferometer to phase changes, assuming a linear relationship.}
    \label{fig:QBER_Phase_TypeII}
\end{figure}

Another important source of quantum bit errors is the phase misalignment of the interferometers.
In \cref{fig:QBER_Phase_TypeII}, we compare measurements of the QBER in the phase basis for varying interferometer phase differences for the \typeII system with simulation results. Both the simulation results and measurements show a cosine-like dependency as expected for an ideal QKD system without imperfections (cf.~\cref{eq:ideal_coincidence_probability}).

\subsection{Key Rates}\label{sec:Key_Rates}

\begin{figure}
    \centering\includegraphics{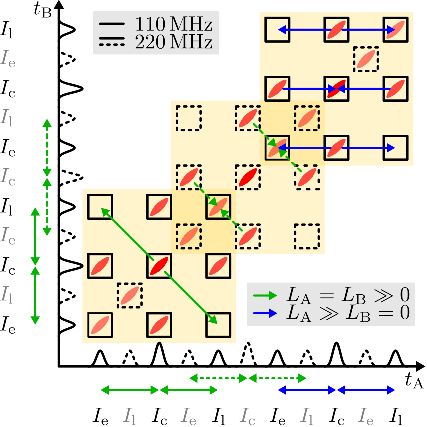}
    \caption{Ordinary ($\SI{110}{\mega\hertz}$) and interleaved ($\SI{220}{\mega\hertz}$) time-bin structure. Chromatic dispersion leads to a broadening along an axis determined by the fiber lengths $L_\upA, L_\upB$, in the direction indicated by the arrows (green: equal fiber lengths, blue: Bob's receiver located close to the source, i.e.\ $L_B =0$).}
    \label{fig:NestedBins}
\end{figure}

To maximize the quantum bit rate, the pump pulse repetition rate $r_\upp$ should be maximized and the pulse width should be minimized. In our setup, the interferometer path lengths are fixed, but we can double the repetition frequency such that the time-bins associated with adjacent pump pulse repetitions are interleaved~\cite{Fitzke_2022_4Party,Dolejsky_2023}. This is depicted in \cref{fig:NestedBins} for the pulse repetition rates of ${r_\upp = \SI{110}{\mega\hertz}}$ and ${r_\upp = \SI{220}{\mega\hertz}}$, where the $\SI{220}{\mega\hertz}$ correspond to the interleaved pulse repetition cycles. The interleaving can lead to overlaps between the intensity of adjacent repetitions which may increase the QBER and thus result in lower secure key rates. The overlap between adjacent repetitions becomes especially relevant when the distance between the source and the participants as well as the bandwidth of the photons are so large that the wave packets are significantly elongated by chromatic dispersion in the transmission fibers. We depict the two special cases where the photon pair source is located in the middle between the receivers, resulting in transmission fibers of equal length and the case where Bob's receiver is placed close to the source.

As all parameters remain invariant over different pump pulse repetitions, cross-talk between adjacent pulses can be modeled by using the fact that the probability of the last repetition causing a detection in the present cycle is the same as the probability of the present repetition causing a detection in the next cycle and vice versa. Consequently, the impact from adjacent repetitions can be determined by computing the detection probabilities of a single pump pulse cycle and considering all affected time-bins, including those that are part of adjacent repetitions. For example, for detecting vacuum in detectors $D_\upA, D_\upB$ and intervals $I_\upA, I_\upB$, the effect of the cross-talk between different repetitions is included by extending the corresponding vacuum POVM expectation to
\begin{equation}
    \langle \hat{\Uppi}_{I_\upA,\vac}^{(D_\upA)} \hat{\Uppi}_{I_\upB,\vac}^{(D_\upB)} \rangle \to \prod_{n=-\infty}^\infty \underbrace{\langle \hat{\Uppi}_{I_\upA+n/r_\upp,\vac}^{(D_\upA)} \hat{\Uppi}_{I_\upB+n/r_\upp,\vac}^{(D_\upB)} \rangle}_{\text{with } r_{\mathrm{noise}}^{(D_\upA)} \,=\, r_{\mathrm{noise}}^{(D_\upB)} \,=\, 0 \text{ for } n \,\neq\, 0} \,.
\end{equation}
Here, ${I+n/r_\upp}$ is the interval obtained by shifting the interval $I$ by $n$ multiples of the repetition time ${1/r_\upp}$. For the shifted intervals the noise rate is set to zero as the noise is already included in the original interval.

In the simulation, for large distances chromatic dispersion may lead to numerical difficulties due to fast oscillations arising from the chirped signal. To circumvent this, we use the method proposed in~\refcite{Andrianov_2016} to isolate the fast oscillations from the slower change of the envelope. In terms where the oscillation frequency exceeds a certain threshold, the integral values average to zero in good approximation and do not need to be resolved. 

\begin{figure}
    \includegraphics{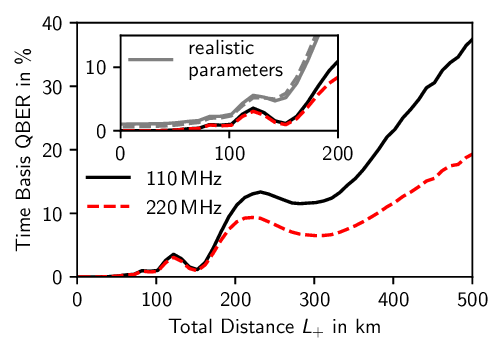}
    \caption{\TypeII time basis QBER when varying the distance $L_+$ between Alice and Bob assuming equal lengths ${L_\upA = L_\upB}$, ideal detectors and no losses for ${\mu = 0.02, r_\upp = \SI{110}{\mega\hertz}}$ and ${\mu = 0.01, r_\upp = \SI{220}{\mega\hertz}}$. In this setting, the phase QBER is largely unaffected by the chromatic dispersion while the time QBER shows oscillations caused by the side lobes in the JSA (cf.in \cref{fig:TypeII_JSD}).}
    \label{fig:QBER_Dispersion_TypeII}
\end{figure}

We investigated the effect of cross-talk between time-bins due to chromatic dispersion by simulating the time basis QBER. \Cref{fig:QBER_Dispersion_TypeII} shows the results for a \typeII setup without losses, noise clicks and dead times to isolate the effects from chromatic dispersion.
The total distance ${L_+ = L_A + L_B}$ for ${L_A = L_B}$ between Alice and Bob is varied for the repetition rates of ${\SI{110}{\mega\hertz}}$ and ${\SI{220}{\mega\hertz}}$. The inset shows a comparison to the simulation results obtained when using the mean values of the parameter ranges determined for the real setup.

The chromatic dispersion stretches the JSA in ${t_- = t_\upA - t_\upB}$ direction due to the equal fiber lengths as depicted in \cref{fig:NestedBins}~(green arrows), leading to an increase of the time basis QBER for longer distances, as the broadening causes the JSA to extend into the early-late/late-early time-bin combinations. However, the time basis QBER does not exhibit a monotonic behaviour but shows oscillations. 
Those occur because the chromatic dispersion maps the side lobes of the photon pair spectrum (cf.~\cref{fig:TypeII_JSD}) into the time domain. With an increasing fiber length, the side lobes begin to extend into time-bin combinations corresponding to quantum bit errors.

For the ideal setup, the higher repetition frequency yields a lower time QBER. This might seem counter-intuitive at first, because the time-bins are more densely packed, such that the chromatic dispersion becomes relevant already for shorter fiber lengths. However, the reason for the counter-intuitive dependency becomes apparent when considering the structure of the time-bins in \cref{fig:NestedBins} in more detail. The time-bin interleaving leads to photons in time-bin combinations corresponding to basis mismatches, i.e.\ early/late time-bin for one party and central time-bin for the other, leaking into the time-bins corresponding to early-early and late-late combinations of adjacent repetitions. Thus, additional correct events in intervals representing the time basis appear, which are not present at the lower repetition rate.

\begin{figure*}
    \subfloat[\Typezero system]{
        \includegraphics{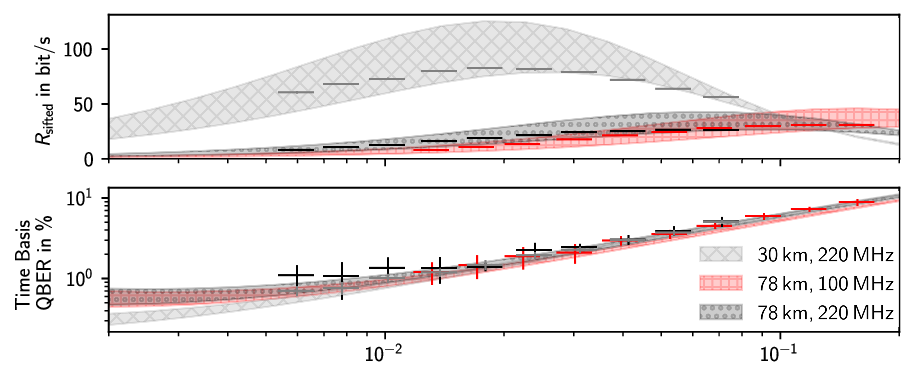}
        \label{fig:Keyrates_QBER_Type0}
    }\\
    \subfloat[\TypeII system.]{
        \includegraphics{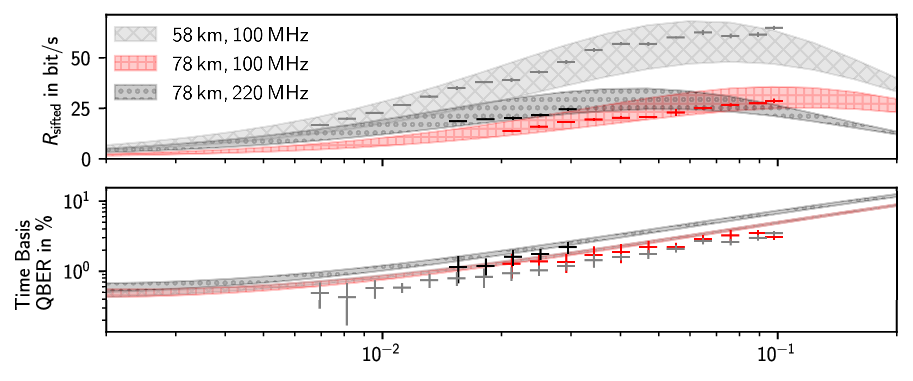}
        \label{fig:Keyrates_QBER_TypeII}
    }
    \caption{Comparison of the sifted key rate $R_{\text{sifted}}$ and time basis QBER between simulated (error bands) and measured (error bars) data when varying the mean number of generated photon pairs $\mu$. The error bands indicate the values between the best- and worst-case simulation scenarios, representing the uncertainties of the measured system parameters. The vertical error bars indicate the statistical uncertainty of the measurements and the horizontal error bars indicate the experimental uncertainty of the mean number of photon pairs. Measurements and simulations have been performed for fiber lengths of ${(L_\upA, L_\upB) \in \{ (\SI{27.3}{\kilo\meter}, \SI{30.8}{\kilo\meter}), (\SI{27.3}{\kilo\meter}, \SI{50.4}{\kilo\meter})\}}$ in \cref{fig:Keyrates_QBER_Type0} and for fiber lengths of ${(L_\upA, L_\upB) \in \{ (\SI{27.3}{\kilo\meter}, \SI{50.4}{\kilo\meter}), (\SI{9.6}{\kilo\meter}, \SI{20.5}{\kilo\meter})\}}$ in \cref{fig:Keyrates_QBER_TypeII}, respectively.}
    \label{fig:Keyrates_QBER}
\end{figure*}

Finally, we examine the effect of multi-pair creations on the key generation by varying the mean number of photon pairs $\mu$ in the simulation of both setups. In the experiment, the QBER in the phase basis is subject to major statistical fluctuations caused by thermal fluctuations and the algorithm that keeps the interferometer phases aligned with each other. Therefore, here we only present the QBER in the time basis and refer to \cref{fig:QBER_Phase_TypeII} for the results on the phase basis QBER. As stated above, due to only two participants using the whole \typeII SPDC spectrum, the mean number of generated photon pairs per pulse is sufficiently small such that the corresponding squeezing parameters allow us to employ the Poissonian approximation. The resulting analytical expressions required to evaluate the detection probabilities are derived in \cref{sec:Appendix_Analytical_Expressions_for_the_Detection_Probabilities} and are given by \cref{eq:FinalCovariance_Trace,eq:FinalCovariance_SquareTrace}.
\Cref{fig:Keyrates_QBER_Type0} and \cref{fig:Keyrates_QBER_TypeII} show a comparison between measured and simulated values of the sifted key rates and time basis QBERs for the \typezero system and the \typeII system, respectively.

For both systems, two different values of the pump pulse repetition rate $r_\upp$, as well as two different distances between the participants have been considered. The larger repetition rate of ${r_\upp = \SI{220}{\mega\hertz}}$ corresponds to the interlacing of two time-bin structures as presented in \cref{fig:NestedBins}. For small values of the mean number of photon pairs generated per pump pulse $\mu$, we observe an increase in the sifted key rate as expected. At some point, however, the behaviour of the sifted key rate is dominated by detector saturation due to the dead time of the SPADs. This is due to the individual detection rates becoming so high, that the probability of all four detectors between the two parties being active decreases so far that the sifted key rates eventually start to decrease when further increasing $\mu$. Simultaneously, the QBER increases in $\mu$ due to an increasing number of multi-photon-pair generations and afterpulses. For the \typeII setup, due to the broad spectrum, we can observe a significant impact of chromatic dispersion for the nested case, as the pulses become sufficiently close to each other to trigger detections in adjacent time-bins and increase the QBER (cf.~\cref{fig:QBER_Dispersion_TypeII}). All of the effects discussed above are well-represented in both the simulations and the measurement data.

\FloatBarrier
\section{Summary and Conclusions}

We presented an in-depth time-/frequency-resolved simulation and analysis of an entanglement-based quantum key distribution system employing the time-bin encoded BBM92 protocol. This serves as an example to demonstrate and validate the expansions derived in the first part of this series, \refcite{Kleinpass_2024_partI}. These expansions allow us to extend the covariance formalism of Gaussian states to the temporal and spectral degrees of freedom while keeping the requirements for numerical performance low, even for highly entangled quantum states. Taking the spectrum of the generated photon pairs as well as the frequency-dependent losses they exhibit into account is important for such simulations as they directly affect the Franson interference visibility and hence the performance of the QKD systems.

Two different realizations of the system have been examined, one of which employs \typeII spontaneous parametric down-conversion to generate entangled photons, for which we applied the Poisson approximation presented in \cite{Kleinpass_2024_partI} and derived the resulting analytic expressions for the detection probabilities here. The other realization employs a \typezero process for which we developed a method to restrict the joint spectral amplitude to an effective interval around the channels of the wavelength division multiplexer used in that setup to greatly restrict the numerical effort required to evaluate the resulting expressions for the detection probabilities of interest.

We showed that our approach is well-suited to incorporate a number of experimental imperfections directly into the simulations by representing them in the covariance formalism. This includes the generation of multiple photon pairs, chromatic dispersion in the transmission fibers and mode mismatches in the receiver interferometers. We demonstrated how measured photon spectra deviating from the ideal theoretical sinc\textsuperscript{2} shape can be integrated by reconstructing the complex spectrum from the measured power spectrum and presented a detailed model of the photon detection in the different time bins, taking into account dark counts, afterpulses and the dead times of the detectors. 

We thoroughly characterized all devices in the setup to obtain a set of realistic parameters describing the system and compared the simulated key rates and quantum bit error rates to corresponding measurements. The values closely match, validating the ability to reliably predict the system's performance over a multitude of different parameter settings.

For a given distance between the parties and the source, there will be an optimal set of parameters including the mean number of generated photon pairs, the pulse repetition rate and the detector settings leading to the maximum secure key generation rate for given post-processing parameters. Performing simulation like the one presented here allows to accurately examine and predict the performance of such systems for various scenarios, enabling an efficient optimization of the system parameters and the identification of individual factors limiting the performance to subsequently improve them.

Future work could apply the methods we presented to other kinds of quantum optical systems employing biphoton states. For example, polarization-entangled QKD systems could be modeled by introducing  polarization as an additional discrete degree of freedom for each spatial mode or by considering photon generation in non-linear media allowing for multiple spatial modes. 

\appendix
\FloatBarrier

\section{Analytical Expressions for the Detection Probabilities} \label{sec:Appendix_Analytical_Expressions_for_the_Detection_Probabilities}
\begin{table}
    \centering
     \caption{Coefficients $K_{x,y}^{(D)}$ in \cref{eq:sdagger_P_s}. All quantities implicitly carry another label $\upA,\upB$ distinguishing between Alice and Bob, e.g.\ ${K_{\upA,\ups,\upl}^{(0)} = \eta_{\upA,0}^{(\ups)} \eta_{\upA,0}^{(\upl)} \xi_\upA^2 \mathcal{T}_\upA^2 \mathcal{R}_\upA^2}$.}
    \label{tab:PTC_Coefficients}
\begin{tabular}{cc@{\hskip 5pt}|@{\hskip 5pt}c@{\hskip 10pt}c@{\hskip 10pt}c}
        \toprule
        \multicolumn{2}{c@{\hskip 5pt}|@{\hskip 5pt}}{\multirow{2}{*}{$K_{x,y}^{(D)}$}} & \multicolumn{3}{c@{\hskip 5pt}}{Path combination $(x,y$)}\\[3pt]
        & & $(\ups,\ups)$ & $(\ups,\upl), (\upl,\ups)$ & $(\upl,\upl)$\\
        \midrule
        \multirow{2}{*}{\rotatebox[origin=c]{90}{Detec.}} & $0$ & $\eta_0^{(\ups)} \eta_0^{(\ups)} \mathcal{T}^4$ & $\eta_0^{(\ups)} \eta_0^{(\upl)} \xi^2 \mathcal{T}^2 \mathcal{R}^2$ & $\eta_0^{(\upl)} \eta_0^{(\upl)} \mathcal{R}^4$\\[3pt]
        & $1$ & $\eta_1^{(\ups)} \eta_1^{(\ups)} \mathcal{T}^2 \mathcal{R}^2$ & $-\eta_1^{(\ups)} \eta_1^{(\upl)} \xi^2 \mathcal{T}^2 \mathcal{R}^2$ & $\eta_1^{(\upl)} \eta_1^{(\upl)} \mathcal{T}^2 \mathcal{R}^2$\\
        \bottomrule
    \end{tabular}
\end{table}
This section presents the procedure of obtaining the transformation matrices for the QKD setup analytically and then translating the resulting expressions into the continuous-mode formalism. The starting point is the renormalized covariance after the SPDC, obtained by inserting \cref{eq:Covariance_Type0TypeII} into \cref{eq:RenormalizedCovariance_Definition}, as this is the last active transformation applied to the system. The remaining passive transformations are given by \cref{eq:RenormalizedCovariance_TransformationBBM92}. For the \typezero setup, the wavelength channels are sufficiently far away from the central frequency for the probability of both photons entering the same channel to vanish, i.e. \cref{eq:WDM_Condition_NoDoublePhotons} is fulfilled. For the \typeII system a perfect polarizing beam splitter is assumed, such that both photons end up in separate spatial modes. Then, according to \cref{eq:Covariance_Type0_WDM} and \cref{eq:Covariance_TypeII}, the renormalized covariance after the SPDC takes the form ${\bm{\varGamma} = \bar{\bm{\varGamma}} \oplus \cc}$, where, on the discrete-mode level, $\bar{\bm{\varGamma}}$ is a ${2 \times 2}$ matrix corresponding to Alice and Bob's DOFs. The interferometer and mode mismatch beam splitters introduce 10~additional discrete modes, which are initially all in the vacuum state. In total this yields 12 discrete modes: Each of the four detectors receives one interfering signal mode and two non-interfering mismatch modes. As both photons will never interact again, the remaining transformations take the form ${\bm{S} = \bm{S}_0 \bm{S}_\omega = (\bar{\bm{S}}_0 \bar{\bm{S}}_\omega) \oplus \cc}$ as well (cf.~\cref{eq:RenormalizedCovariance_TransformationBBM92}). The  distinction between $\bm{S}_0$ and $\bm{S}_\omega$ is made because applying $\bm{S}_0$ can only lead to linear time shifts in the Fourier domain which can be accounted for analytically, while applying $\bm{S}_\omega$ may introduce non-trivial effects such as temporal elongation of the wave packets due to chromatic dispersion. The final renormalized covariance takes the form ${\bm{\varGamma}_{\mathrm{final}} = \bar{\bm{\varGamma}}_{\mathrm{final}} \oplus \cc}$, where ${\bar{\bm{\varGamma}}_{\mathrm{final}} = \bar{\bm{S}}_0 (\bar{\bm{\varGamma}}_\omega \oplus \bm{0}_{10 \times 10}) \bar{\bm{S}}_0^\dag = \bar{\bm{s}} \bar{\bm{\varGamma}}_\omega \bar{\bm{s}}^\dag}$ and ${\bar{\bm{\varGamma}}_\omega = \bar{\bm{s}}_\omega \bar{\bm{\varGamma}} \bar{\bm{s}}_\omega^\dag}$. The reduced transformation takes the form ${\bar{\bm{s}} = \bar{\bm{s}}_\upA \oplus \bar{\bm{s}}_\upB^*}$ and can be computed by symbolic matrix multiplication at the discrete-mode level, yielding \cref{eq:ReducedTransformation_BBM92}.

For modeling time-resolved detection events, the expression is transformed to the time domain by applying the symplectic Fourier transformation $\bm{F}$ to obtain ${\tilde{\bm{\varGamma}}_{\mathrm{final}} = \bm{P} \tilde{\bm{s}}^\dag \tilde{\bm{\varGamma}} \tilde{\bm{s}} \bm{P}}$, where
\begin{IEEEeqnarray}{c"c"c}\label{eq:ReducedTransform_TimeCovariance}
        \tilde{\bm{s}} = \bm{F} \bar{\bm{s}} \bm{F}^\dag &
        \text{and} &
    \tilde{\bm{\varGamma}} = \bm{F} \bar{\bm{\varGamma}}_\omega \bm{F}^\dag \,.
\end{IEEEeqnarray}
Using \cref{eq:DeterminantSylvester} yields the final expression for the vacuum detection probability ${P_{\vac} = \abs{\det(\mathds{1} + \tilde{\bm{s}}^\dag \bm{P} \tilde{\bm{s}} \tilde{\bm{\varGamma}})}^{-1}}$. On the discrete-mode level, by arranging the expressions in this way, the dimension of the final renormalized covariance is effectively reduced from ${24 \times 24}$ to ${2 \times 2}$.\nolinebreak\footnote{Note that ${\tilde{\bm{s}}^\dag \bm{P} \tilde{\bm{s}} \tilde{\bm{\varGamma}}}$ does not fulfill the properties of a (renormalized) covariance operator, because it is not Hermitian. However, under application of the (Fredholm) determinant in \cref{eq:FredholmDeterminant_LogarithmExpansion} it is equivalent to the renormalized covariance ${\bm{P} \tilde{\bm{s}} \tilde{\bm{\varGamma}} \tilde{\bm{s}}^\dag \bm{P}}$, effectively reducing its dimension whenever the detection probabilities are computed.} It is not necessary to evaluate the total transformation $\bm{S}$, which, in the naive approach, is obtained by many ${24 \times 24}$ matrix multiplications. Instead, $\bar{\bm{s}}$ can be directly computed from the same amount of (symbolic) ${12 \times 2}$ matrix multiplications, which decreases the computational cost of the implementation significantly. Afterwards, the resulting expressions are translated into the continuous-mode formalism by identifying the different symbols with the actions of the corresponding integral operators.

As stated in \cref{eq:RenormalizedCovariance_PumpIF}, the renormalized covariance can be written as a sum over the contributions of the short (s) and long (l) paths ${z \in \{ \ups, \upl\}}$ of the pump interferometer:
\begin{widetext}
    \begin{equation}
        \tilde{\bm{\varGamma}}_z(t, t') = \frac{1}{2}
        \mqty{
            \bm{u}_z(t) \mleft[ \mathfrak{c}_N(\bm{\sigma}_z) - \Id \mright] \bm{u}_z^\dag(t') &
            \upe^{\upi \phi_{\upp,0}^{(z)}} \bm{u}_z(t) \mathfrak{s}_N(\bm{\sigma}_z) \bm{v}_z^\dag(t')\\
            \upe^{-\upi \phi_{\upp,0}^{(z)}} \bm{v}_z(t) \mathfrak{s}_N(\bm{\sigma}_z) \bm{u}_z^\dag(t') &
            \bm{v}_z(t) \mleft[ \mathfrak{c}_N(\bm{\sigma}_z) - \Id \mright] \bm{v}_z^\dag(t')
        }
    \end{equation}
    for $z \in \{\ups, \upl\}$, where
    \begin{subequations}
        \begin{IEEEeqnarray}{rCl}
            \bm{u}_z(t_\upA) &=& \bm{\mathcal{F}}_{\omega_\upA}\mleft( \upe^{\upi \phi_\upA^{(\mathrm{CD})}(\omega_\upA)} \bm{u}_\upA(\omega_\upA) \mright)(t_\upA-\tau_\upp^{(z)}) \,, \IEEEeqnarraynumspace\\
            \bm{v}_z(t_\upB) &=& \bm{\mathcal{F}}_{\omega_\upB}\mleft( \upe^{\upi \phi_\upB^{(\mathrm{CD})}(\omega_\upB)} \bm{v}_\upB(\omega_\upB) \mright)(t_\upB-\tau_\upp^{(z)}) \,.
            \IEEEeqnarraynumspace
        \end{IEEEeqnarray}
    \end{subequations}
    Here, $\bm{u}_\upA$ and $\bm{v}_\upB$ are vectors obtained by applying the frequency-dependent transmission functions of the channels to the Schmidt modes of the original JSA and $\phi_\rho^{(\mathrm{CD})}$ for ${\rho \in \{ \upA, \upB \}}$ are the quadratic phases acquired due to chromatic dispersion in the optical fibers to Alice and Bob.
    The reduced transformation in \cref{eq:ReducedTransformation_BBM92} only contains frequency-independent terms and linear phases, so that the total transformation including the projection onto the detectors and time intervals of interest that acts onto the renormalized covariance is directly obtained:
    \begin{equation}\label{eq:sdagger_P_s}
        (\tilde{\bm{s}}_\rho^\dag \bm{P}_\rho \tilde{\bm{s}}_\rho)(t, t') = \sum_{x,y \in \{\ups, \upl\}} \sum_{D\in\{0,1\}}
        K_{\rho,x,y}^{(D)} \rect_I^{(D)}(t+\tau_\upA^{(x)}) \upe^{\upi (\phi_{\rho,0}^{(y)} - \phi_{\rho,0}^{(x)})} \delta(t-t'+\tau_\rho^{(y)}-\tau_\rho^{(x)}) \,,
    \end{equation}
    with the coefficients $K_{x,y}^{(D)}$ given in \cref{tab:PTC_Coefficients}. Thereby, the final expression after applying all transformations to the renormalized covariance reads
    \begin{IEEEeqnarray}{rCl}\label{eq:FinalCovariance_BBM92}
        (\tilde{\bm{s}}^\dag \bm{P} \tilde{\bm{s}} \tilde{\bm{\varGamma}})(t, t') &=& \frac{1}{2} \sum_{x, y, z \in \{\ups, \upl\}} \sum_{D \in \{0, 1\}}
        \mqty{
            K_{\upA,x,y}^{(D)} \rect_I^{(D)}(t + \tau_\upA^{(x)}) \upe^{\upi (\phi_{\upA,0}^{(y)} - \phi_{\upA,0}^{(x)})} & 0\\
            0 & K_{\upB,x,y}^{(D)} \rect_I^{(D)}(t + \tau_\upB^{(x)}) \upe^{\upi (\phi_{\upB,0}^{(y)} - \phi_{\upB,0}^{(x)})}
        } \IEEEnonumber\\
        &&\times
        \mqty{
            \bm{u}_z(t + \tau_\upA^{(x)} - \tau_\upA^{(y)}) \mleft[ \mathfrak{c}_N(\bm{\sigma}_z) - \Id \mright] \bm{u}_z^\dag(t') &
            \upe^{\upi \phi_{\upp,0}^{(z)}} \bm{u}_z(t + \tau_\upA^{(x)} - \tau_\upA^{(y)}) \mathfrak{s}_N(\bm{\sigma}_z) \bm{v}_z^\dag(t')\\
            \upe^{-\upi \phi_{\upp,0}^{(z)}} \bm{v}_z(t + \tau_\upB^{(x)} - \tau_\upB^{(y)}) \mathfrak{s}_N(\bm{\sigma}_z) \bm{u}_z^\dag(t') &
            \bm{v}_z(t + \tau_\upB^{(x)} - \tau_\upB^{(y)}) \mleft[ \mathfrak{c}_N(\bm{\sigma}_z) - \Id \mright] \bm{v}_z^\dag(t')
        } \,.
        \IEEEeqnarraynumspace
    \end{IEEEeqnarray}
When employing the lowest-order approximation of the covariance expansion up to ${N = 2}$ in \cref{eq:RenormalizedCovariance_SeriesExpansion}, it is more convenient to write this expression without using the Schmidt decomposition of the JSA. For a \typeII process, in the absence of frequency-dependent channel transmissions, \cref{eq:FinalCovariance_BBM92} becomes    
\begin{IEEEeqnarray}{rCl}\label{eq:FinalCovariance_LowestOrder}
        (\tilde{\bm{s}}^\dag \bm{P} \tilde{\bm{s}} \tilde{\bm{\varGamma}})(t, t') &=& \sum_{x,y,z \in \{\ups, \upl\}} \sum_{D \in \{0,1\}} 
        \mqty{
            K_{\upA,x,y}^{(D)} \rect_I^{(D)}(t + \tau_\upA^{(x)}) \upe^{\upi (\phi_{\upA,0}^{(y)} - \phi_{\upA,0}^{(x)})} & 0\\
            0 & K_{\upB,x,y}^{(D)} \rect_I^{(D)}(t + \tau_\upB^{(x)}) \upe^{\upi (\phi_{\upB,0}^{(y)} - \phi_{\upB,0}^{(x)})}
        } \IEEEnonumber\\
        && \times
        \mqty{
            \mu K_{\upp,z}^2 (\bm{\psi}_z \bm{\psi}_z^\dag)(t + \tau_\upA^{(x)} - \tau_\upA^{(y)}, t') &
            \sqrt{\mu} K_{\upp,z} \upe^{\upi \phi_{\upp,0}^{(z)}} \bm{\psi}_z(t + \tau_\upA^{(x)} - \tau_\upA^{(y)}, t')\\
            \sqrt{\mu} K_{\upp,z} \upe^{-\upi \phi_{\upp,0}^{(z)}} \bm{\psi}_z^\dag(t + \tau_\upB^{(x)} - \tau_\upB^{(y)}, t') &
            \mu K_{\upp,z}^2 (\bm{\psi}_z^\dag \bm{\psi}_z)(t_\upB + t_\upB^{(x)} - t_\upB^{(y)}, t')
        } \,,
        \IEEEeqnarraynumspace
    \end{IEEEeqnarray}
    where
    \begin{IEEEeqnarray}{c"c"c}
        \psi_z(t_\upA, t_\upB) = \psi_{\mathrm{CD}}(t_\upA, t_\upB - \tau_\upp^{(z)}) &
        \text{with} &
        \psi_{\mathrm{CD}}(t_\upA, t_\upB) = \mathcal{F}_{\omega_\upA,\omega_\upB}\mleft( \upe^{\upi (\phi_\upA^{(\mathrm{CD})} + \phi_\upB^{(\mathrm{CD})})} \psi(\omega_\upA, \omega_\upB) \mright)(t_\upA, t_\upB) \,.
        \IEEEeqnarraynumspace
    \end{IEEEeqnarray}
    Applying the second order approximation of the logarithm expansion and neglecting terms of $\mathcal{O}(\psi^4)$, i.e.\ employing the Poissonian approximation, yields
    \begin{IEEEeqnarray}{Rcl}\label{eq:FinalCovariance_Trace}
        \Tr \big( \tilde{\bm{\varGamma}}_{\mathrm{final}} \big) = \mu \sum_{\mathclap{x,y,z, D}} K_{\upp,z}^2 \Big( && K_{\upA,x,y}^{(D)} \upe^{\upi (\phi_{\upA,0}^{(y)}-\phi_{\upA,0}^{(x)})} \int_{I_\upA^{(D)}} \dl t_\upA \int \dl t\, \psi_{\mathrm{CD}}(t_\upA - \tau_\upp^{(z)} - \tau_\upA^{(y)}, t) \psi_{\mathrm{CD}}^*(t_\upA - \tau_\upp^{(z)} - \tau_\upA^{(x)}, t) \IEEEnonumber\\
        + && K_{\upB,x,y}^{(D)} \upe^{\upi (\phi_{\upB,0}^{(x)} - \phi_{\upB,0}^{(y)})} \int_{I_\upB^{(D)}} \dl t_\upB \int \dl t\, \psi_{\mathrm{CD}}^*(t, t_\upB - \tau_\upp^{(z)} - \tau_\upB^{(y)}) \psi_{\mathrm{CD}}(t, t_\upB - \tau_\upp^{(z)} - \tau_\upB^{(x)}) \Big) \,, \IEEEnonumber\\
    \end{IEEEeqnarray}
    describing the single-photon interference and
    \begin{IEEEeqnarray}{Rcl}\label{eq:FinalCovariance_SquareTrace}
        \norm{\tilde{\bm{\varGamma}}_{\mathrm{final}}}_{\mathrm{HS}}^2 = 2\mu \sum_{\mathclap{\substack{x, y, z, D\\x', y', z', D'}}} 
        K_{\upp,z} K_{\upp,z'} K_{\upA,x,y}^{(D)} K_{\upB,x',y'}^{(D')} && \upe^{\upi (\phi_{\upp,0}^{(z)} - \phi_{\upp,0}^{(z')} + \phi_{\upA,0}^{(y)} - \phi_{\upA,0}^{(x)} + \phi_{\upB,0}^{(y')} - \phi_{\upB,0}^{(x')})} \IEEEnonumber\\
        \int_{I_\upA^{(D)}} \dl t_\upA \int_{I_\upB^{(D')}} \dl t_\upB\, && \psi_{\mathrm{CD}}(t_\upA - \tau_\upp^{(z)} - \tau_\upA^{(y)}, t_\upB - \tau_\upp^{(z)} - \tau_\upB^{(\mu')}) \IEEEnonumber\\
        \times && \psi_{\mathrm{CD}}^*(t_\upA - \tau_\upp^{(z')} - \tau_\upA^{(x)}, t_\upB - \tau_\upp^{(z')} - \tau_\upB^{(x')})+ \mathcal{O}(\psi^4)\,,
        \IEEEeqnarraynumspace
    \end{IEEEeqnarray}
\end{widetext}
corresponding to two-photon interference. Including higher orders of photon-pair generations requires the evaluation of $\mathcal{O}(\psi^4)$ integrals, which, in our approach, would allow us to employ the bivariate Hermite approximation discussed in \refcite{Kleinpass_2024_partI}.

\section*{Acknowledgements}

This research has been funded by the Deutsche Forschungsgemeinschaft (DFG, German Research Foundation), under Grant No.\ SFB 1119--236615297. We thank Till Dolejsky for contributing the QKD experiments.

\bibliography{bibliography}% Produces the bibliography via BibTeX.
\end{document}